%% file: serpjet_accepted.tex
\documentclass[structabstract]{aa}  
\usepackage{graphicx}
\usepackage{lscape}
\usepackage{txfonts}
\usepackage{natbib}
\bibpunct{(}{)}{;}{a}{}{,}
\begin{document}
\bibliographystyle{aa}

   \title{Proper motions of embedded protostellar jets in Serpens\thanks{Based on observations made with the Nordic Optical Telescope, operated on the island of La Palma jointly by Denmark, Finland, Iceland, Norway, and Sweden, in the Spanish Observatorio del Roque de los Muchachos of the Instituto de Astrofisica de Canarias. }}


   \author{A.A. Djupvik
          \inst{1}
          \and
           T. Liimets
          \inst{2,3}
          \and
          H. Zinnecker
          \inst{4,5}
          \and
          A. Barzdis
          \inst{6}
          \and
           E. A. Rastorgueva-Foi
          \inst{7}
          \and
           L. R. Petersen
          \inst{8}
          }

   \institute{Nordic Optical Telescope, Rambla Jos\'{e} Ana Fern\'{a}ndez P\'{e}rez, 7,
              ES-38711 Bre\~{n}a Baja, Spain \\
              \email{amanda@not.iac.es}
         \and
             Tartu Observatory, T\~oravere, 61602, Estonia \\
             \email{tiina@obs.ee}
         \and
             Institute of Physics, University of Tartu, Ravila 14c, 50411, Estonia
         \and
             SOFIA Science Center, NASA Ames Research Center, 94035 Moffett Field, USA \\
             \email{hzinnecker@sofia.usra.edu}
         \and
             Deutsches SOFIA Institut (DSI), Univ. Suttgart, Germany
         \and 
             Institute of Astronomy, University of Latvia, Raina bulv. 19, Riga, LV 1586, Latvia
          \and
             School of Maths \& Physics, University of Tasmania, Australia, Hobart TAS 7001 \\
            \email{efoi@utas.edu.au}
          \and
             Niels Bohr Institute, University of Copenhagen, Juliane Maries Vej 30, DK-2100 
             Copenhagen, Denmark
             }

\date{Received May 14, 2015; accepted December 18, 2015}

 
  \abstract
   {To investigate the dynamical properties of protostellar jets.}
   {Determine the proper motion of protostellar jets around Class\,0 and
    Class\,I sources in an active star forming region in Serpens.}
   {Multi-epoch deep images in the 2.122 $\mu$m line of shocked molecular 
    hydrogen, v=1-0 S(1), obtained with the near-infrared instrument 
    NOTCam over a time-scale of 10 years, are used to determine proper 
    motion of knots and jets. K-band spectroscopy of the brighter knots is
    used to supply radial velocities, estimate extinction, excitation
    temperature, and H$_2$ column densities towards these knots. }
   {We measure the proper motion of 31 knots over different time scales
    (2, 4, 6, and 10 years). The typical tangential velocity is around 50 
    km/s for the 10 year base-line, but for shorter time-scales a maximum 
    tangential velocity up to 300 km/s is found for a few knots. 
    Based on morphology, velocity information and the locations of known 
    protostars, we argue for the existence of at least three partly 
    overlapping and deeply embedded flows, one Class\,0 flow and two
    Class\,I flows. The multi-epoch proper motion results indicate time-variable 
    velocities of the knots, for the first time directly measured for a Class\,0 
    jet. We find in general higher velocities for the Class\,0 jet than for the 
    two Class\,I jets. While the bolometric luminosites of the three driving 
    sources are about equal, the derived mass flow rate \.{M}$_{\rm out}$ is two 
    orders of magnitude higher in the Class\,0 flow than in the two Class\,I flows. 
    }
    {}

   \maketitle
%

\section{Introduction}

Protostellar jets are spectacular manifestations of the birth of a star.  Supersonic and 
collimated jets drive bipolar outflows of molecular gas, removing angular momentum from 
-- and allowing accretion onto the protostar. The origin of the jets is poorly understood, 
but a tight relation between the accretion disk and the jet is evident in magneto-hydro-dynamic 
models. For a recent review on jets and outflows see \citet{fra14}.
The power carried by the jet is expected to be proportional to the accretion luminosity of the 
central star, suggesting that protostellar jets from the youngest protostars have higher 
velocities. As shown by \citet{bont96} the youngest Class\,0 sources (ages from a few 10$^3$ to 
a few 10$^4$ yrs) have more energetic CO outflows than the more evolved Class\,I protostars. 
Studies of shocked molecular hydrogen in the jets have shown an empirical relation between 
the H$_2$ jet luminosity and the bolometric luminosity of the driving source, but found no clear 
distinction between Class\,0 and Class\,I sources \citep{car06}.

Plasma is ejected in the form of a jet which interacts with the circumstellar medium via 
radiative shocks, giving rise to observable emission in various lines over a large spectral 
range \citep{rei01,bac11}. The near-infrared emission from such shocks is mainly in 
lines of [Fe~II] and the ro-vibrational lines of molecular hydrogen. The observed features 
of the latter are referred to as Molecular Hydrogen emission-line Objects (MHO's). In 
particular, the v=1-0~S(1) $\mathrm{H}_2$ line at 2.122~$\mu$m is a good tracer of shock excited 
emission in low-velocity shocks, and the 2 $\mu$m region suffers less from extinction than 
optical lines and makes it possible to observe the most embedded flows, potentially the very 
youngest jets. 

Protostellar jets are often complex morphological structures, composed of a multitude of 
knots and features, and the school-book example of a symmetric bi-polar jet is rarely observed.
In regions of dense star formation one frequently finds asymmetric flows, jets without 
counter-jets, variable ejection, colliding knots, precession, or two jets that are interacting. 
Mapping the kinematics of the individual knots will facilitate identification of whole entities 
of jets and make it possible to separate individual flows in crowded regions. This is required 
in order to study the kinematic and physical properties of the jets and their driving sources.

Two-epoch proper motions will give the average tangential velocities between the two epochs, 
while multi-epoch imaging will allow us to investigate how velocities may vary. Recent 
multi-epoch ($>$ 2 epochs) measurements in the optical as well as the near-IR of nearby jets in 
Taurus and Chamaeleon have revealed time-variable proper motions and even interaction among 
knots \citep{bon08,car09}. 
In order to determine the velocities of protostellar jets, we need to identify the 
individual flows and their driving sources. To do this we are in the need of both proper motion 
and radial velocity information.

In this paper we study the dynamics of the individual knots of embedded protostellar jets in the 
Serpens NH3 cloud, also referred to as the Serpens/G3-G6 region or Serpens Cluster B (see 
\citet{eir08} for a review of the region). These jets were first presented in \citet{dju06} 
together with new protostar candidates 
discovered from a near-IR, mid-IR, submm and radio study of the region. Based on morphology
only we tentatively identified the driving sources to be the new Class\,0 candidate MMS3 for the 
North-South oriented jet, and a double Class\,I for the NE-SW oriented jet. MMS3 was confirmed by 
\citet{eno09} in their extensive protostar survey (where it is named Ser-emb-1) to be an {\it early} 
Class\,0 of very low temperature. While being presumably the youngest protostar in Serpens, it has 
already developed a protostellar disk \citep{eno11}. Studying these jets allows for a comparison 
between Class\,0 and Class\,I jets in the same region.

Due to the large cloud extinction to this embedded cluster, these jets are not detected in 
the optical. No extended emission is found in deep H$\alpha$ imaging except for the faint 
Herbig-Haro object HH~476, discovered in an optical [SII] survey by \citet{zie99}, and some 
emission around the association of bright TTauri stars named Serpens/G3-G6 \citep{coh79}. 

We obtained deep $\mathrm{H}_2$ line imaging in four epochs over 10 years. The first epoch of 
data (2003) was already published in \citet{dju06}, and the second epoch (2009) was taken during the 
Nordic-Baltic Summer School in Turku, Finland.\footnote{Nordic-Baltic Optical/NIR and Radio Astronomy 
Summer School: ``Star Formation in the Milky Way and Nearby Galaxies'', 8-18 June 2009, Turku, 
Finland. See URL http://www.not.iac.es/tuorla2009/local/} Preliminary results, carried out as a 
student project during the course showed that proper motions indeed could be measured between 2003 
and 2009. The data is re-reduced here, and in addition, we obtained two more epochs (2011 and 2013), 
as well as K-band spectroscopy of selected knots, all obtained with NOTCam at the Nordic Optical 
Telescope. 

Section~\ref{tar-obs-red} describes the target field, observations and data reduction, 
Sect.~\ref{idflow} presents the proper motion, fluxes, and radial velocity measurements of the
knots, as well as the identification of three distinct flows and their velocities, while 
Sect.~\ref{physical} describes the derivation of physical parameters for the three flows, and 
Sect.~\ref{con} gives our summary and conclusions. 

\begin{figure}[t]
\centering
\includegraphics[width=9cm]{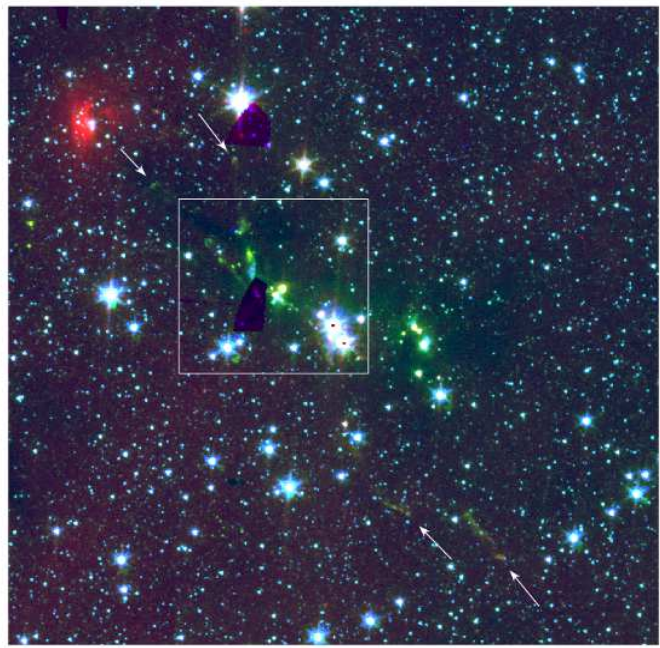}
\caption{
Overview of the Serpens/G3-G6 region (also called Serpens Cluster B and the Serpens NH3 
cloud) as seen by IRAC/Spitzer in 3.6 $\mu$m (blue), 4.5 $\mu$m (green), and 8 $\mu$m (red). 
North up, East left. Total FOV $\approx 15' \times 15'$. The jets are mainly enhanced in the 4.5 
$\mu$m band. The $\approx 4'$ field monitored with NOTCam for proper motions is marked white and 
shown in Fig.~\ref{h2image}. The white arrows point to two bow-shock shaped features 
outside our monitored field to the North and North-East and two jet-like
features in the South-West region. }
\label{irac}
\end{figure}

\begin{figure}
\centering
\includegraphics[width=9.3cm]{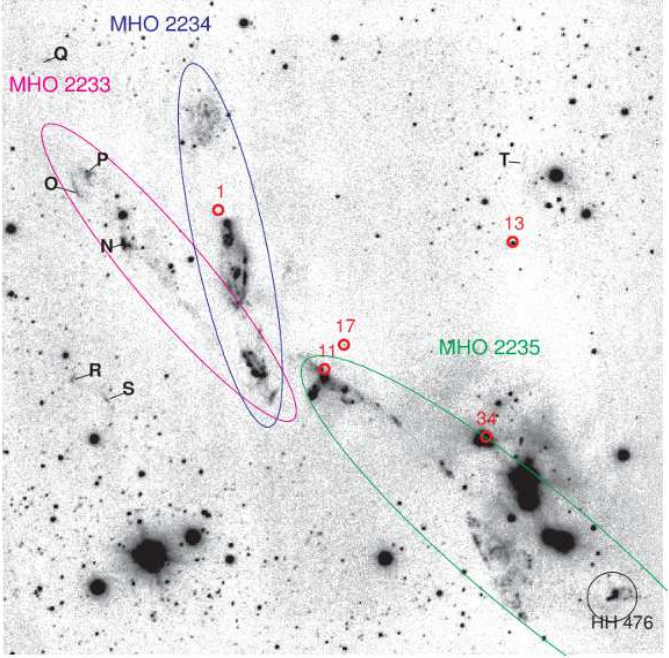}
\caption{Deep H$_2$ + continuum (2.122 $\mu$m) narrowband NOTCam image of the 
$\approx 4'$ field observed in this multi-epoch study. North up, East left.
Naming of jet structures according to the MHO Catalogue is sketched: MHO~2233 
(magenta),  MHO~2234 (blue), and MHO~2235 (green). The previously known HH~476
is marked. Red circles show the positions of Class\,0 and Class\,I protostars, 
the ID number according to designations by \citet{eno09}, for instance 
{\bf 1} refers to the Class\,0 source Ser-emb-1, which is named is MMS3 in 
\citet{dju06}. Only a few knots are identified with letters here, 
and the majority are identified in zoomed-in regions in the following figures. }
\label{h2image}
\end{figure}


\section{Target field, observations, and data reductions}
\label{tar-obs-red}

\subsection{Target field}

The Spitzer/IRAC image in Fig.~\ref{irac} showing the 3.6 $\mu$m (blue), 4.5 $\mu$m 
(green) and 8 $\mu$m (red) bands gives an overview of the Serpens NH3 region, also 
called Serpens Cluster B. It is centred on a group of four optically visible and very 
bright TTauri stars Ser/G3-G6 \citep{coh79}. Two NH3 cores are located to either side, 
one to the NE and another to the SW, where active star formation takes place and a 
number of Class\,0 and Class\,I objects are found \citep{dju06,har06,har07}. The multi-epoch 
H$_2$ line imaging with NOTCam was obtained in the region outlined by the white box. The 
jets stand out in the 4.5 $\mu$m band emission (green), and apparently they extend 
beyond the field covered by NOTCam. 
In Fig.~\ref{h2image} the jets are outlined by the H$_2$ line emission at 2.122~$\mu$m. 
With a narrow-band continuum image centred at 2.087~$\mu$m obtained in 2003 the line 
emission knots were easily distinguished from the continuum \citep{dju06}. In the Molecular 
Hydrogen Emission-Line Objects (MHO) catalogue \citep{dav10} the objects studied in this 
paper are numbered \object{MHO~2233}, \object{MHO~2234}, and \object{MHO~2235}, their 
positions being sketched in Fig.~\ref{h2image}. Judging from the Spitzer image, it 
seems that MHO~2233 extends further to the North-East, and possibly MHO~2235 
to the South-West. MHO~2234 extends further to the North, and our 2009 observations 
confirm 2.122 $\mu$m emission in the northernmost bow-shaped feature seen in the IRAC 
image in Fig.~\ref{irac}, but we have only one epoch of data for it. 

Each of the MHO objects are composed of a number of smaller knots which we identify and 
name according to the guidelines established in \citet{dav10}. It is not clear, however, 
from the 2D morphology which knot belong to which jet and how many jets there are. Based 
on proper motions and radial velocities we show in the following that we find three jets 
driven by three protostars. 

\subsection{Imaging}
\label{ima}

The Nordic Optical Telescope's near-IR Camera/Spectrograph, NOTCam\footnote{NOTCam is 
documented in detail on the URL http://www.not.iac.es/instruments/notcam/.}, was used 
to obtain deep multi-epoch narrow-band images in the H$_2$ v=1-0~S(1) line at 2.122 $\mu$m 
(NOT filter \#218) in order to measure proper motions. Four epochs, hereafter referred to 
as ep1, ep2, ep3, and ep4, are sampled at the years 2003, 2009, 2011, and 2013. The four 
deep images were obtained with the same instrument, but ep1 was imaged with the NOTCam 
engineering grade array, while the science array was used for the other three epochs. The 
wide-field camera of NOTCam has a pixel scale of $0.234''$/pix and a $4' \times 4'$ field 
of view. The deep imaging of the extended emission was mostly performed in beam-switch 
mode to correctly estimate the sky. In these cases the template NOTCam observing script 
``notcam.beamswitch'' was used to obtain a 9-point grid with small-step dithering both ON 
target and in the OFF field, going alternatingly between the ON and OFF regions. The OFF 
field was chosen either to the East, the West or the North of the central field, the West
field being the optimal choice, as it contained no bright disturbing stars whose 
persistency would affect the background estimate. In the North field we found an additional 
bow-shaped feature emission line object in 2009. In the 2003 epoch no beamswitch was applied, 
only small-step dithering around the target field. All narrow-band images were obtained with 
the ramp-sampling readout mode, where the detector is read a number of times during the 
integration, typically using 60s integrations with 10 equally spaced readouts. A linear 
regression analysis is performed pixel by pixel through all reads to obtain the final image.

Table~1 lists the various observations and filters used. The filter ID numbers are according 
to the internal NOT id numbers, where 218 refers to the filter centred on the v=1-0~S(1)
line at 2.122 $\mu$m, while 210 and 230 refer to narrow-band filters representing the 
continuum emission at 2.267 and 2.087 $\mu$m, respectively. For all filters differential
twilight flats were obtained every night. For flux calibration two standard star fields were 
observed before and after the target on the photometric night of May 25th 2013, our reference 
epoch. 

\begin{table}
\caption{Observing log for NOTCam imaging used in this paper.
}
\label{table:1}    
\centering                       
\begin{tabular}{l l c l}        
\hline\hline            
Date        & Filters  & Epoch & Total exptime (sec) \\   
\hline                     
28 May 2003 & 218,230  & ep1   & 2000, 2000 \\
29 May 2007  & J,H,Ks,218,210 &   -       & 450, 450, 450, 900, 900 \\
10 May 2009 & 218      &  ep2   & 540  \\
11 Jun 2009 & 218      &  ep2   & 2100 \\
13 Aug 2011\tablefootmark{1} & 218        &  ep3   & 2700  \\ 
25 May 2013 & 218      &  ep4   & 2430 \\
\hline                                          
\end{tabular}

\tablefoottext{1}{This epoch does not cover the eastern part of the field.} \\
\end{table}

The images were reduced using IRAF\footnote{IRAF is distributed by the National
Optical Astronomy Observatory, which is operated by the Association of Universities
for Research in Astronomy (AURA) under cooperative agreement with the National Science 
Foundation.} and a set of scripts in an external IRAF package notcam.cl v2.6, made for 
NOTCam image reductions. All individual raw images were corrected for zero valued bad 
pixels and hot pixels by interpolation and then divided by the master flats obtained 
from typically 8 differential twilight flats per filter per night. By using differential 
flats, the thermal and dark emission is eliminated from the flats. The individual target 
frames were then sky subtracted using the off fields obtained in beamswitch mode or the 
dithered target images themselves. After this step all individual images were corrected 
for the internal camera distortion. For data from 2009 and onwards we use the distortion 
model for the K$_S$-band available on the NOTCam calibration web page. For the 2003 data 
set, a different distortion model was developed based on data from the same night. After 
distortion correction, all dithered images were registered and combined to form one deep 
image per epoch. 

For flux calibration we used the zeropoint offset between the K$_S$ and the H$_2$ line 
filter (\#218) found from 8 standard stars in photometric conditions in epoch 4 to be 
2.56~$\pm$~0.01 magnitudes, giving a zeropoint for the H$_2$ line filter of 19.94 mag 
(for 1 ADU/sec, Vega magnitudes). Assuming the standards have equal flux at 2.12 and 2.14 
(K$_S$) $\mu$m, we use the relation for a zero'th magnitude star at 2.121 $\mu$m: 
F$_{\lambda}$ = $4.57 \times 10^{-10}$ W m$^{-2}$ $\mu$m$^{-1}$ \citep{tok05}, and the FWHM of 
the filter (0.032 $\mu$m), to arrive at a flux conversion of 
$1.54 \times 10^{-16}$ erg s$^{-1}$ cm$^{-2}$ for 1 ADU/sec. The standard deviation of the 
background in the 2350s deep image of epoch 4 is $1.5 \times 10^{-16}$ erg s$^{-1}$ cm$^{-2}$ 
arcsec$^{-2}$, and the brightest knot, MHO~2235-B, peaks at $1.65 \times 10^{-14}$ 
erg s$^{-1}$ cm$^{-2}$ arcsec$^{-2}$.

\begin{figure*}[t]
\centering
\includegraphics[width=6cm]{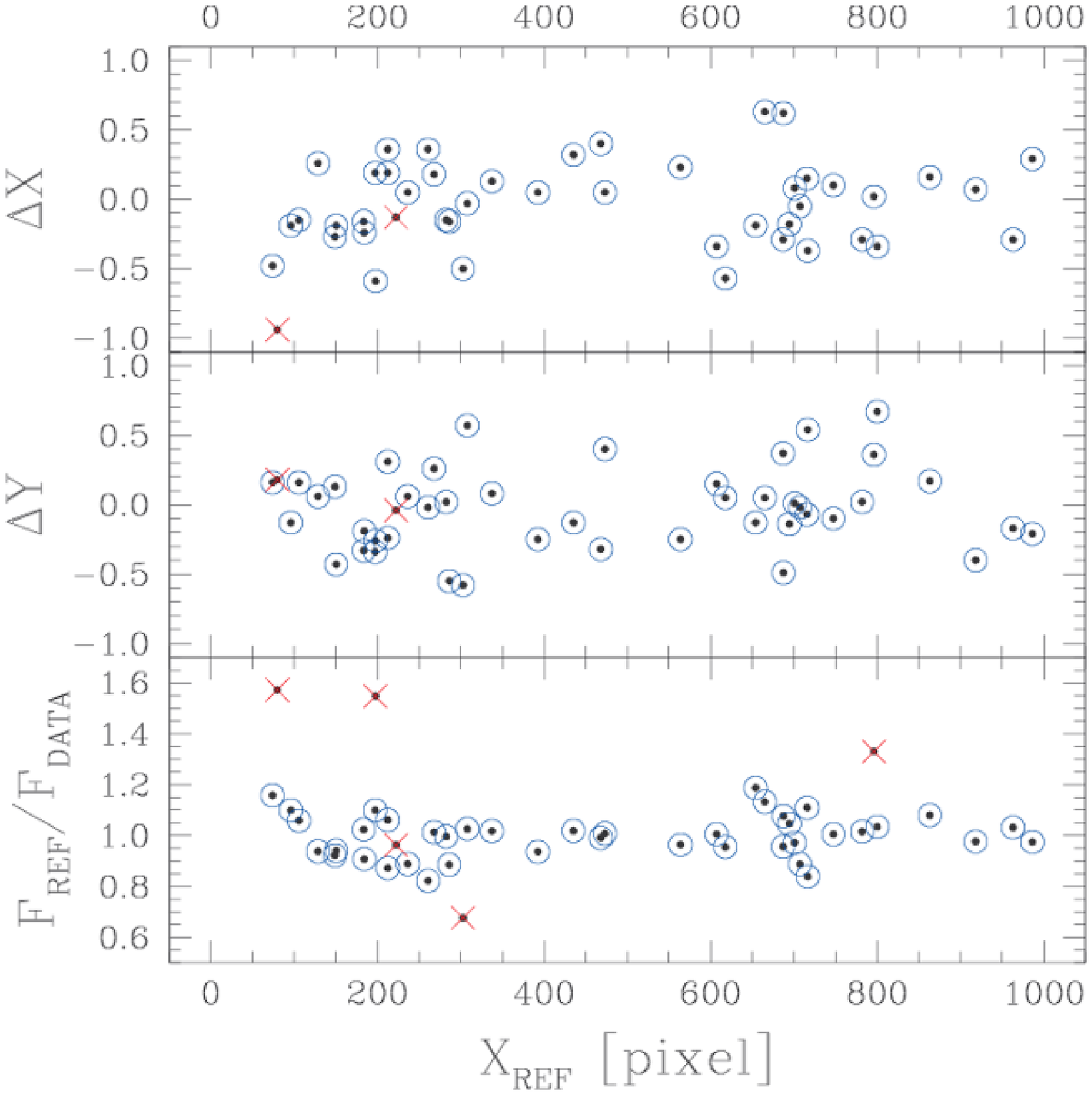}
\includegraphics[width=6cm]{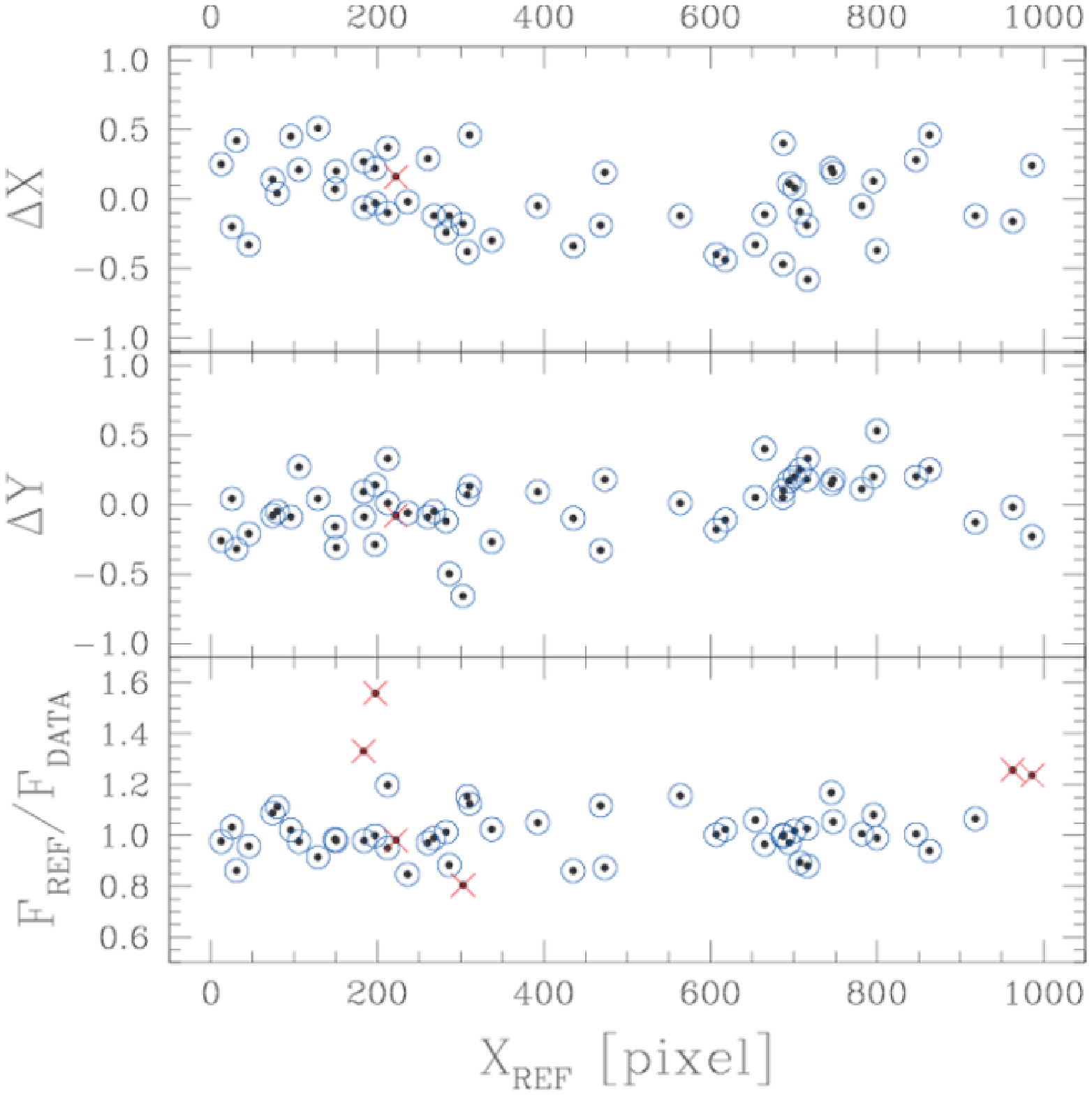}
\includegraphics[width=6cm]{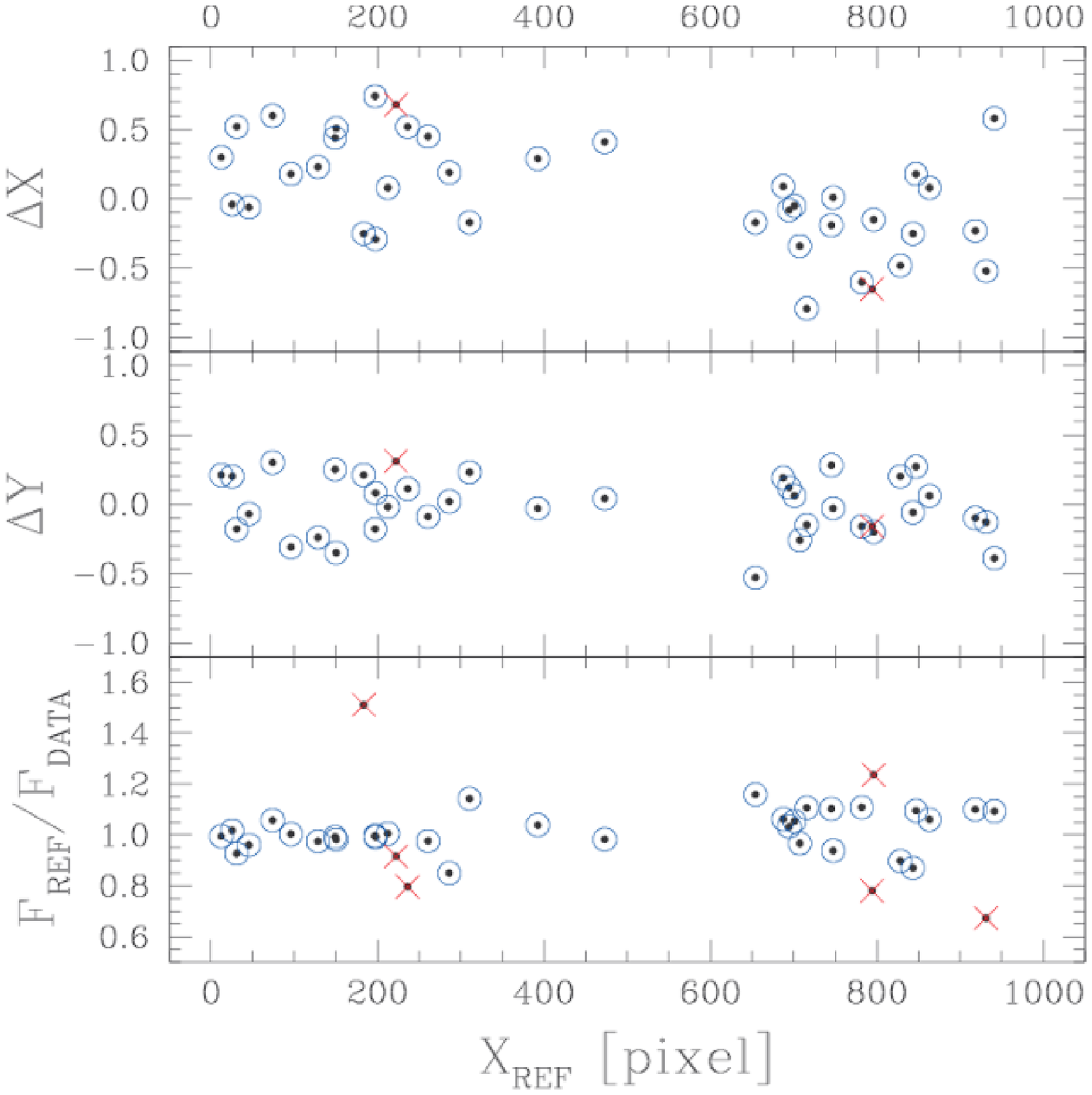}
\caption{
Relative position accuracy between epochs checked on stars after the final 
registration (panels $\Delta X$ and $\Delta Y$) and flux matching accuracy 
checked after seeing and flux matching (lower panels). 
The $\Delta X$ and $\Delta Y$ values are the offsets in X and Y (in pixels) 
between the measured epoch and the reference epoch (i.e. epoch 4). Epoch 1 
is shown to the left, epoch 2 in the centre, and epoch 3 to the right, all 
plotted with respect to reference X-pixel. In the lower panels we show the 
flux ratio for the same stars. Black dots indicate all stars, red crosses 
show stars rejected due to large FWHM and/or high flux ratio, and blue 
circles indicate stars used for the accuracy calculations. 
}
\label{rel_pos_acc}
\end{figure*}

In addition, we obtained broad-band JHKs images to illustrate the extinction in the 
region, the filters and exposure times are listed in Table~\ref{table:1}.
We have also used archive Spitzer data to complement our near-IR imaging with the mid-IR bands 
of IRAC \citep{faz04}. An RGB colour coded IRAC image using the 3.6 $\mu$m (blue), the 4.5 $\mu$m
(green) and the 8 $\mu$m (red) bands is presented in Fig.~\ref{irac}. The jet clearly stands out 
in green by enhancing the cuts for the emission in the 4.5 $\mu$m band, a wavelength region with 
many H$_2$ lines. Note that central parts of both jets have masked out areas in both the 3.6 and 
4.5 $\mu$m bands. The IRAC imaging was obtained in 2004.

\subsection{Spectroscopy}
\label{specobs}

K-band spectroscopy was obtained with NOTCam on October 12th 2011 and March 16th and 17th 
2014, in most cases including several knots in the slit for each pointing. The instrument 
setup was: grism \#1, the 128 $\mu$m slit (a longslit of width $0.6''$ or 2.6 pixels), and 
the K-band filter \#208 used to sort the orders. This covers the range from 1.95 to 2.37 
$\mu$m with a dispersion of 4.1 $\AA $/pix, giving a spectral resolution $\lambda /\Delta 
\lambda$ of about 2100 for the slit used. The slit acquisition was obtained with a previously 
set rotation angle, calculated from the deep images, and using fiducial stars for alignment. 
Before acquiring the target on slit, the exact position of the slit was measured. The spectra 
were obtained by dithering along the slit in an A-B-B-A pattern while auto-guiding the 
telescope. The integration time was 600s per slit position, reading out the array every 60s 
while sampling up the ramp. Argon and halogen lamps were obtained while pointing to target 
in order to minimize effects due to instrument flexure and to improve fringe correction. 
To correct the spectra for atmospheric features we used either HIP92904 (B2 V), HIP91322 (A2 V), 
or HIP92386 (A1 V) as telluric standards and observed them right after the targets. Darks were 
obtained with the same exposure time and readout mode as the target spectra, mainly for 
constructing a hot pixel mask.

For each night hot pixel masks were constructed from darks taken with the same integration time 
and readout mode as the target spectra. Zero-pixels and hot pixels were corrected for by 
interpolation 
in all images. The individual A-B-B-A spectra were sky-subtracted using the nearest offset for 
each, and the 4 sky-subtracted images were aligned, shifted and combined. This worked well for 
target MHO~2235-B where the offset between the A and B components was small, but for MHO~2234, 
where several knots could be aligned in the slit, the A-B offset was large, and each spectrum 
was extracted and wavelength calibrated separately, and then combined. The IRAF task {\it apall} 
was used to extract one-dimensional spectra, and the telluric standard was used as a reference 
for the trace for the target spectra, since these have no continuum emission. For the knots 
MHO~2234-D, MHO~2234-A and MHO~2234-I, the B position spectra resulted too noisy, and only the 
two A-position spectra were combined and used in the analysis.

The target spectra were corrected for telluric features using the IRAF task {\it telluric} after 
interpolating over the stellar absorption lines in the telluric standard star spectra 
(for B2, A1, and A2 dwarfs this concerns Br~$\gamma$ at 2.166 $\mu$m). After division by the 
telluric standard, the spectra were multiplied by a black-body corresponding to the standard 
spectral type to correct the spectral slope. We use the narrow-band imaging of the H2 v=1-0~S(1) 
line from epoch~4 and an effective aperture corresponding to the slit aperture to approximately 
flux calibrate the spectra. The spectral v=1-0~S(1) line flux for MHO~2235-B is thus estimated to 
be $1.4 \times 10^{-14}$ erg s$^{-1}$ cm$^{-2}$.


We took argon and xenon arc lamp exposures while pointing to target, but decided to instead use 
the night sky OH lines at the location of each knot, for a far more precise wavelength calibration. 
Typically 30 lines were used to fit a wavelength solution, using as reference the OH line list by 
\citet{oli13}. Thus, at the slit position of each knot we extracted the OH line spectrum from the 
raw images and made individual wavelength solutions. The typical rms in the fit was 0.3~\AA \ or 
4~km~s$^{-1}$. The instrument profile has a gaussian FWHM of 2.3-2.4 pixels, corresponding to 
10~$\AA $ or 140~km~s$^{-1}$.
 
\begin{landscape}
\begin{table}
\caption{Proper motion measurements of the 31 brightest knots in the jets. The J2000 coordinates 
of the knots are measured in epoch 4 (2013) and are accurate to $\sim 0.06''$. The proper motions 
(PM$_{\rm ij}$) are measured in mas/yr for six different base-lines and listed only for PMs larger 
than 2$\sigma$. The sub-indices {\em ij} give the motion from epoch {\em i} to epoch {\em j}, where 
epoch~1 is 2003, epoch~2 is 2009, epoch~3 is 2011, and epoch~4 is 2013. The proper motion position 
angles (PA$_{\rm ij}$) indicate the direction of the motion from epoch {\em i} to epoch {\em j} and 
the angles are measured from North and Eastwards. 
}
\label{pmtab}     
\input{propermotion.tex}
\end{table}
\end{landscape}

\section{Identifying the flows kinematically}
\label{idflow}

\subsection{Proper motion measurements}
\label{pm} 

We use the four epochs of deep H$_2$ line imaging from year 2003 (ep1), 2009 (ep2), 2011 (ep3) 
and 2013 (ep4) to measure proper motions of the knots.  Particular attention was paid to the image 
registration. The combined deep image from each epoch was registered to the reference frame (ep4) 
using {\it geomap} and {\it geotran} in IRAF with a Legendre polynomial of 3rd order and 30-50 
reference stars. The rms of the fits were: 0.20 pix for ep1, 0.17 pix for 
ep2, and 0.23 pix for ep3. After registering ep1, ep2 and ep3 to the same plate scale as ep4, 
the positional accuracy was again checked by measuring the centroids of all the stars in the field. 
Figure~\ref{rel_pos_acc} shows the individual offsets $\Delta X$ and $\Delta Y$ relative to the 
ep4 X positions. These offsets may include any potential proper motion of the stars, but because 
we use a large number of stars, we find it appropriate to adopt the average standard deviation as 
an estimate of the global positional error. The average of all offsets is 0.28 pixels. 
Excluding the $\Delta X$ of ep3, we get an average of 0.26 pixels or $0.06''$. This value is used 
to derive the uncertainty in our proper motion measurements. 

\begin{figure}
\centering
\includegraphics[width=9cm]{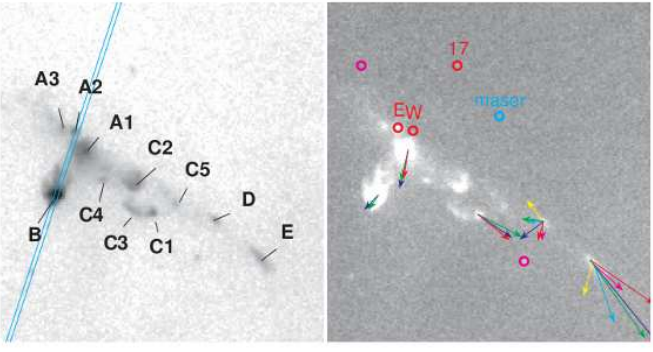}
\caption{The central $50''$ field comprising the MHO~2235 North-Eastern part. North up, East left. 
{\bf Left:} Epoch 4 image overlaid with knot identification and slit position for K-band 
spectroscopy. {\bf Right:} Ep4 image with proper motion vectors, amplified by a factor 20 for 
clarity. The color coding of the pm vectors is as follows: PM$_{14}$ (blue), PM$_{12}$ (red), 
PM$_{13}$ (green), PM$_{24}$ (cyan), PM$_{23}$ (yellow), PM$_{34}$ (magenta).
Protostars are shown with red circles. Ser-emb-11 E and W (a binary Class I) and Ser-emb-17 are 
strong candidate driving sources. Magenta coloured circles denote YSO candidates from \citet{har06}.
The maser refers to an H$_2$O maser position from \citet{bra94}.}
\label{mho2235cen}
\end{figure}

We first tried to define the knots' positions and apertures using contours at a 5$\sigma$ detection 
level, but due to very different intensity levels and diffuse inter-knot emission, we decided not 
to use a general signal-to-noise limit for knot definition but identified them by visual inspection, 
comparing to the deep narrow-band continuum image from 2003, and using the constraint that the peak 
emission is well above 5$\sigma$. In order to determine the knot positions, a Gaussian fit was used 
to find the central X and Y coordinate for each detected knot in every epoch's deep image. 
Measurements were done on the registered images, before seeing and flux matching, in order to keep 
the resolution. For two knots we decided to bin the images in order to derive reliable estimates of 
the knot centres. The accuracy with which we can determine the knot position varies with each knot, 
depending on its actual morphology, but we estimate that this centering error in most cases is much 
smaller than the global positional matching error of 0.26 pixels, which we adopt as our error.

\begin{figure}[t]
\centering
\includegraphics[width=9cm]{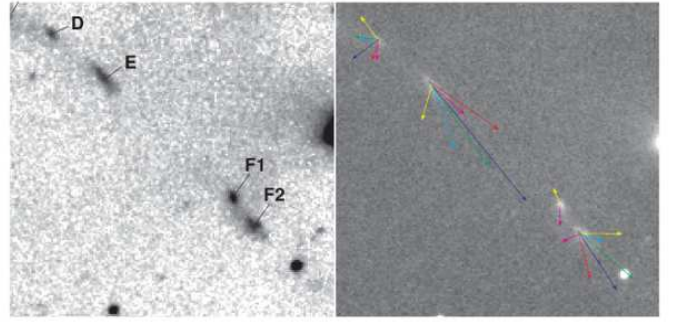}
\caption{The $40''$ field around the fastest moving knot, MHO~2235-E. See text and the caption of 
Fig.~\ref{mho2235cen} for details.
}
\label{mho2235mid}
\end{figure}

\begin{figure}[t]
\centering
\includegraphics[width=9.2cm]{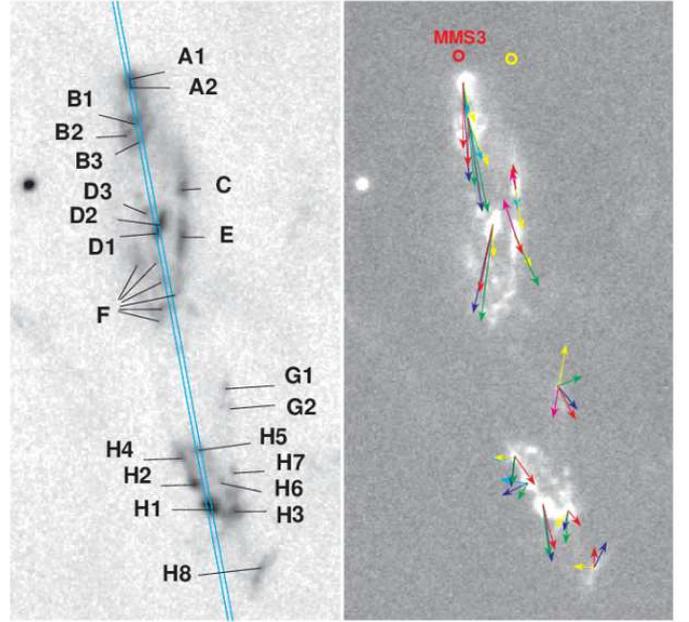}
\caption{The southern part of MHO~2234. See text and the caption of Fig.~\ref{mho2235cen} for 
details. The Class\,0 protostar MMS3 (aka Ser-emb-1, red circle) is probably the driving source. 
The yellow circle shows the location of another red YSO from \citet{har09}.}
\label{mho2234s}
\end{figure}


In total we have identified 57 knots, and for 31 of these we can reliably measure the knot 
positions. These are listed in Table~\ref{pmtab}. The J2000 positions are given for epoch 4 (2013), 
using 46 stars from the 2MASS point source catalogue \citep{skr06} to calculate the plate solution 
with an rms of 0.06'' both in RA and DEC. The proper motions (PM) are calculated in milli arc 
seconds per year (mas/yr) for each of the six base-lines. The length of the PM vectors are given in 
Table~\ref{pmtab} as PM$_{\rm ij}$. The base-lines are indicated by the sub-indices {\em ij}, 
referring to motion from epoch {\em i} to epoch {\em j}. The position angles of the PM vectors are 
given as PA$_{\rm ij}$, measured positively from North and Eastwards, from epoch {\em i} to epoch 
{\em j}. The errors in the proper motion speed (in mas/yr) are calculated for each base-line, using 
the above adopted positional error of 0.26 pixels. The proper motion speed errors are, thus, 
estimated to be: ePM$_{12}$=10, ePM$_{23}$=28, ePM$_{13}$=7, ePM$_{24}$=15, ePM$_{34}$=35, 
and ePM$_{14}$=6 mas yr$^{-1}$. The errors in the position angles of the PM vectors are estimated as 
180/${\rm \pi}$ $\times \,\, {\rm atan2 (ePM/PM)}$. The measurements presented in Table~\ref{pmtab} 
list proper motions whenever larger than 2$\sigma$. We note that knot MHO~2234-A1 has emerged over 
the course of our observations from being hardly noted in epoch~1 to becoming roughly as bright as 
its close neighbour A2 in epoch~4. No proper motion could be calculated for it.

\begin{figure}[t]
\centering
\includegraphics[width=9.2cm]{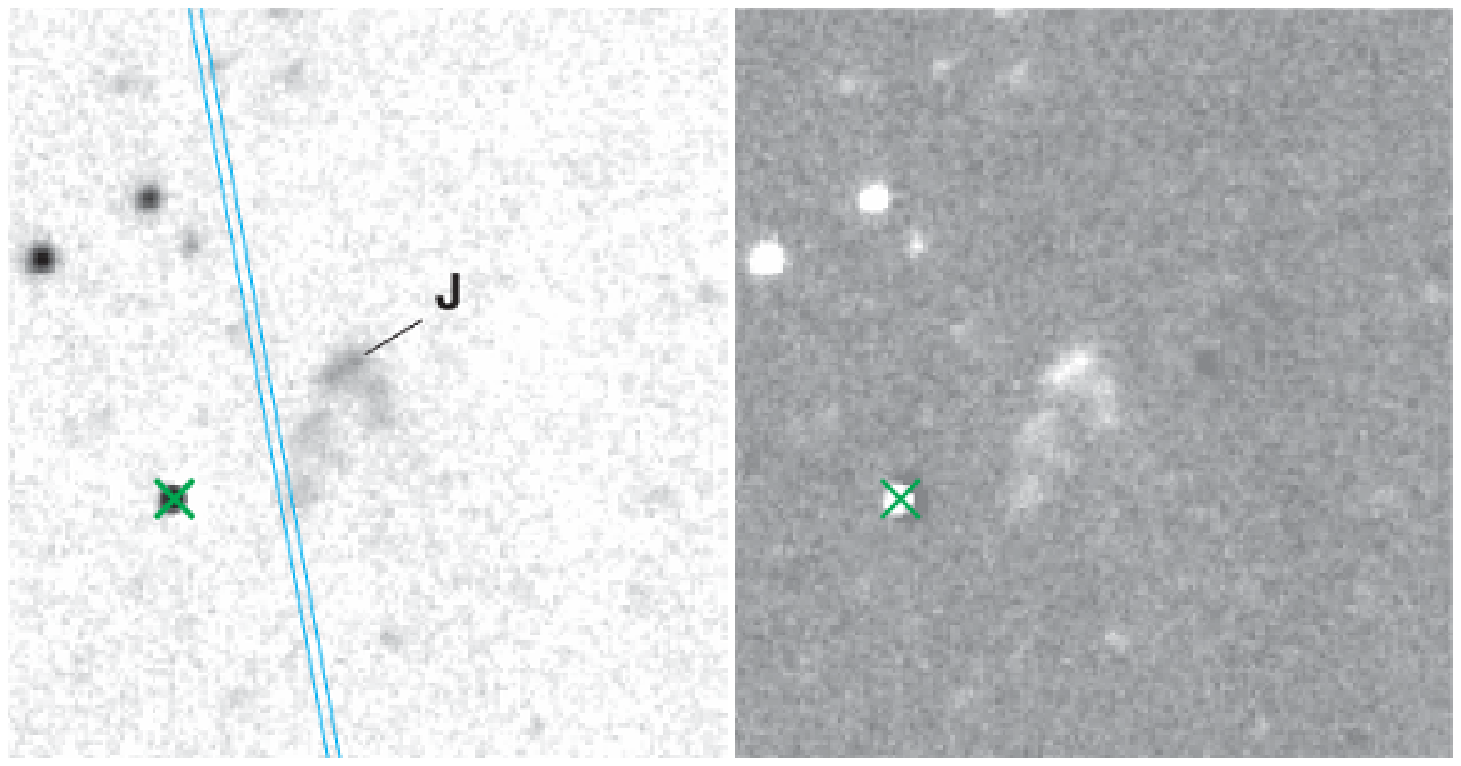}
\includegraphics[width=9.2cm]{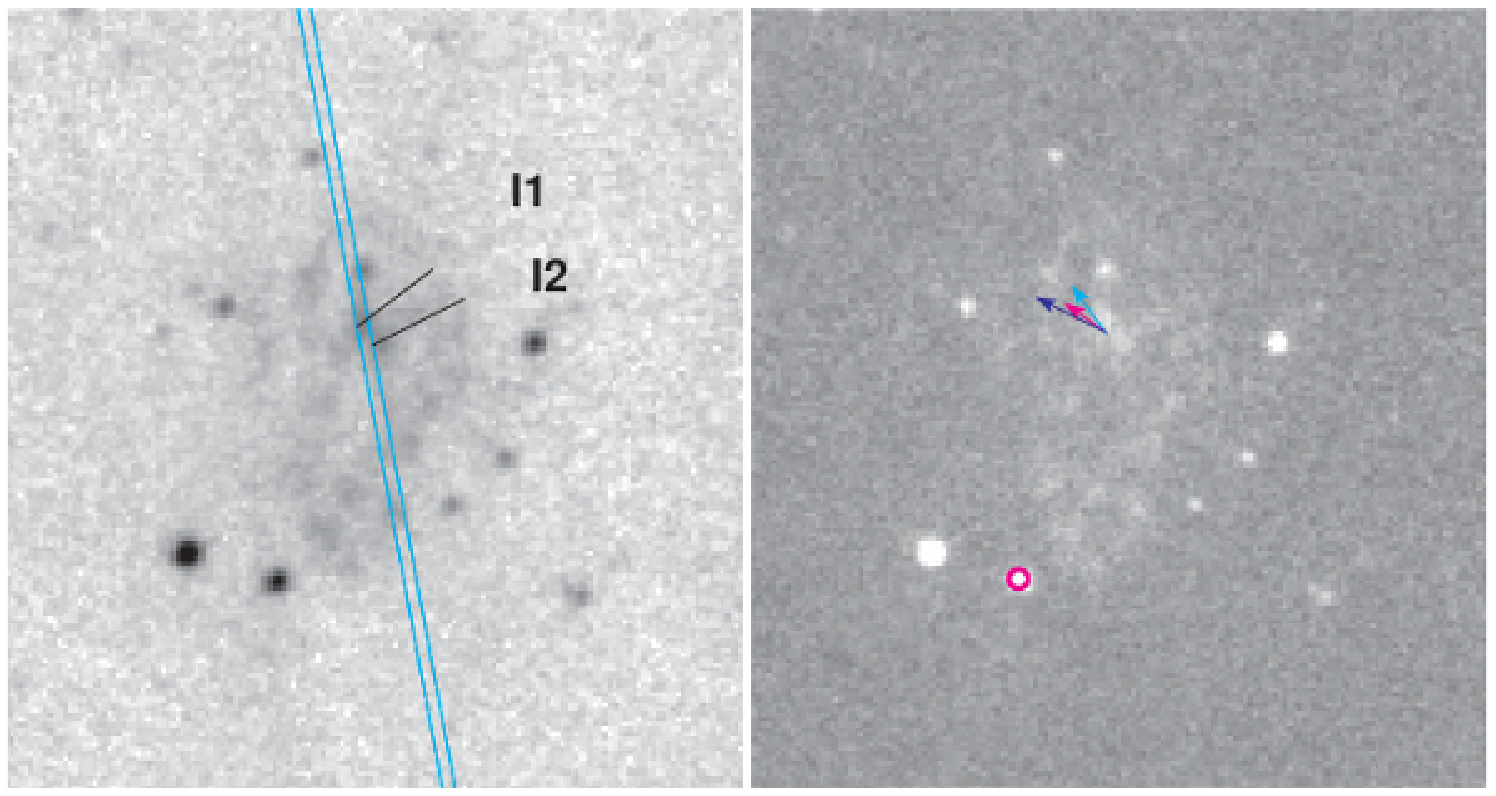}
\caption{The northern parts of MHO~2234, lower panels show the field included in the NOTCam 
monitoring field, while the upper panels show the new knot MHO~2234-J further to the North, 
imaged  only once in 2009. The green cross is an artifact. See text and the caption of 
Fig.~\ref{mho2235cen} for details.}
\label{mho2234n}
\end{figure}

\begin{figure}[t]
\centering
\includegraphics[width=9.2cm]{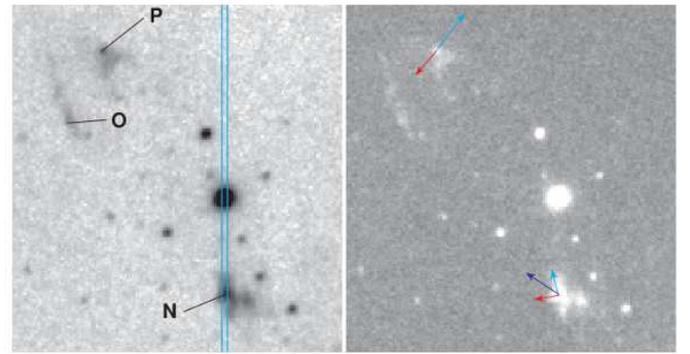}
\caption{The north-eastern part of MHO~2233 ($37''$ field). See text and the caption of 
Fig.~\ref{mho2235cen} for details.
}
\label{mho2233east}
\end{figure}

\begin{figure}[t]
\centering
\includegraphics[width=9.2cm]{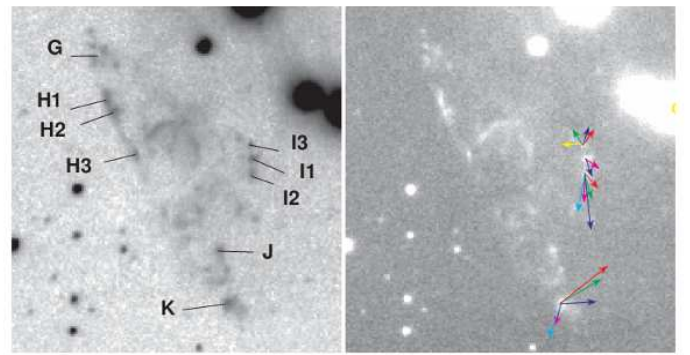}
\caption{The South-Western part of MHO~2235 (47$''$ field). See text and the
caption of Fig.~\ref{mho2235cen} for details.}
\label{mho2235sw}
\end{figure}

\begin{figure}[h]
\centering
\includegraphics[width=9.2cm]{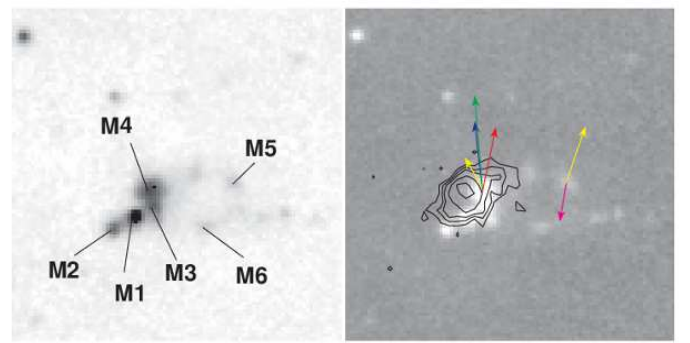}
\caption{Zoom in on the HH~476 region ($20''$ field), the SW part of MHO~2235 in the NOTCam 
field. See text and the caption of Fig.~\ref{mho2235cen} for details. The black contours show 
the H$\alpha$ emission from 2009 of the previously known Herbig-Haro object HH~476.}
\label{hh476}
\end{figure}

The multi-epoch proper motion vectors are overlaid in colours on the zoomed images shown 
in Figs.~\ref{mho2235cen} to \ref{hh476}. Each figure is a zoom-in of a sub-region and 
shows the H$_2$ line (+ continuum) image with knot identification and slit position 
for K-band spectra in the left panel, and the inverse greyscale with the proper motion 
vectors overlaid in the right panel. The PM vector length is amplified by a factor 20 
for clarity, and every base-line is colour coded as explained 
in the caption of Fig.~\ref{mho2235cen}. 

Multiple PM vectors for the same knot follow more or less the same direction (approximately 
within the errors of the position angles), from base-line to base-line, for instance knots 
A1, B, C1 in Fig.~\ref{mho2235cen}, knots A2, B1, D1+D2, H1, in Fig.~\ref{mho2234s}, knot I1 
in Fig.~\ref{mho2234n}, and knot M4 in Fig.~\ref{hh476}. For several knots, however, there 
is a large directional scatter, such as for knots D, E, and F2 in Figs.~\ref{mho2235mid}, 
and knots G1, H3, H4, H8, in Fig.~\ref{mho2234s}. In these cases the maximum spread of the 
multi-epoch vectors for one and the same knot can be 90 degrees or more. Most of these are 
relatively faint or elongated knots, and it could be that we have underestimated the positional 
uncertainty due to a more difficult centering. Thus, it is unclear whether this scatter is 
real or merely reflects a larger measurement error for these knots. 

\begin{figure}[t]
\centering
\includegraphics[width=9.2cm]{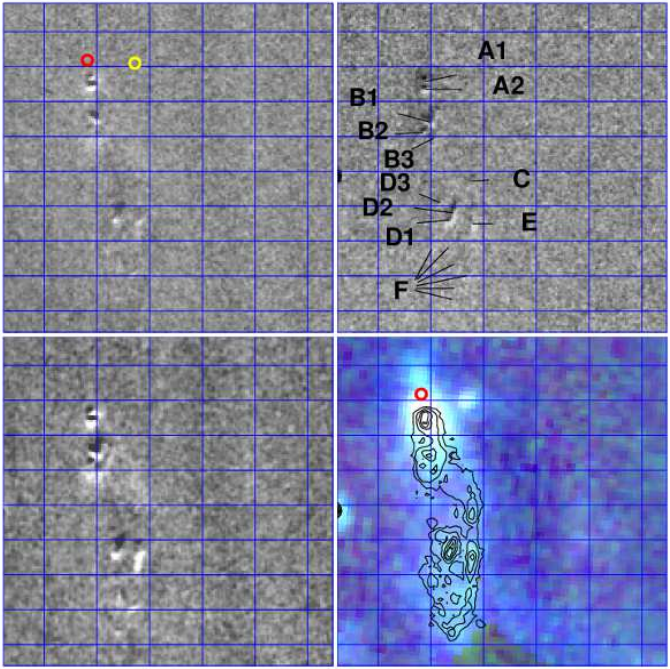}
\caption{Difference images for part of the MHO~2234 flow including the position of the 
Class\,0 source MMS3 or Ser-emb-1 (red circle) and a not classified YSO from \citet{har09}
(yellow circle). The image shifts are in the sense: ep2-ep1 (upper left), ep3-ep2 (upper right), 
and ep4-ep1 (lower left), bright regions showing positive flux. The IRAC/Spitzer 
4.5 $\mu$m (blue), 5.8 $\mu$m (green) and 8 $\mu$m (red) is shown with the H$_2$ v=1-0 S(1) line 
emission in black contours (lower right).
}
\label{imshift}
\end{figure}

\begin{figure}[t]
\centering
\includegraphics[width=9.2cm]{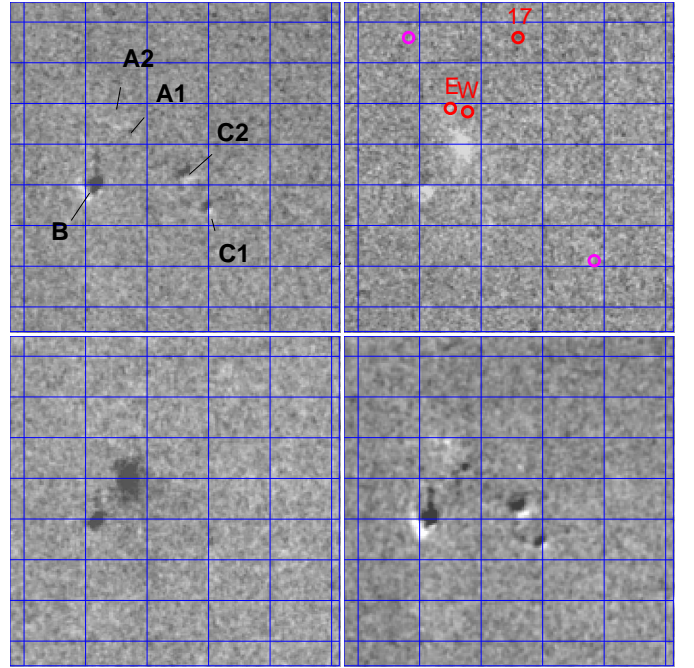}
\caption{Difference images for part of the MHO~2235 flow including the position of the Class~I 
sources Ser-emb-17 and Ser-emb-11 E and W (red circles) and two YSO candidates from \citet{har06}
(magenta circles), cf Fig.~\ref{mho2235cen}.
The image shifts are in the sense: ep2-ep1 (upper left), ep3-ep2 (upper right), ep4-ep3 
(lower left) and ep4-ep1 (lower right), bright regions showing positive flux. The brightening 
near knot A1 appeared in the 2011 image and had disappeared again in the 2013 image.}
\label{imshift2}
\end{figure}

In addition to the PM measurements we made difference images between epochs using the seeing 
matched and flux matched images. Figure~\ref{imshift} shows part of the southern arm of the 
MHO~2234 jet, including knots A1, A2, B1, B2, B3, C, D1, D2, E and the F knots (cf. 
Fig.~\ref{mho2234s}), and Fig.~\ref{imshift2} shows the same region as that in 
Fig.~\ref{mho2235cen}. When a knot has moved it shows up as a negative shadow next to a 
positive peak of emission, although, difference images are also highly sensitive to slight 
changes in the knot morphology and/or flux. Figure~\ref{imshift2} demonstrates the brightening 
around knot MHO~2235-A1 that appeared some time between 2009 and 2011 and disappeared again 
some time before the 2013 observations, a clear example of rapid changes in protostellar jets.
In general, there is agreement between the PM vectors and the shifts in the difference images.

\subsection{Tangential velocities}

There has been a discrepancy in the distance estimates to Serpens between the VLBI parallax 
measurement of EC95 giving a distance of 415 $\pm$ 5~pc for the Serpens Main Cluster 
\citep{dzi10}, a neighbouring cloud located $< 1^{\circ}$ to the north, and extinction studies 
looking at the global properties of the Serpens--Aquila cloud complex which find smaller 
distances; 225 $\pm$ 55 pc according to \citet{str03} and 203 $\pm$ 7 pc according to 
\citet{knu11}. The extinction map of the Serpens-Aquila region shown in \citet{bont10} and 
the discussions in \citet{mau11} suggest that the Northern part of the complex, comprising 
Serpens Main and NH3, is a separate entity from the rest of the complex, seen in projection, 
and located at a larger distance. In this paper we will follow this reasoning and adopt 
415~pc as the distance. For the longest base-line of about 10 years, the uncertainty in the 
tangential velocity is 12 km~s$^{-1}$ while for the shortest base-line of less than 2 years, 
it is 67 km~s$^{-1}$, see Table~\ref{stat}. The highest velocity measured over the longest 
base-line is 190~km~s$^{-1}$ for knot MHO~2235-E, see Fig.~\ref{mho2235mid}. This scales down 
to roughly half the value if we apply the 220 $\pm$ 80 pc distance estimate \citep{str03}. 

\begin{figure}[t]
\centering
\includegraphics[angle=270,width=9cm]{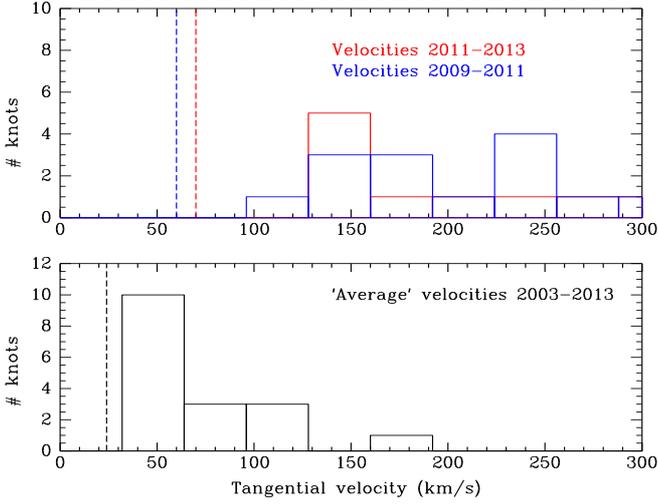}
\caption{Number distribution of the measured tangential velocities (in km~s$^{-1}$, assuming 
d=415 pc). The average velocities over a 10 year timespan from 2003 to 2013 (lower panel) are 
compared to the more ``instantaneous'' velocities measured over the shortest timescales of about 
2 years (upper panel), the 2011-2013 period (bold red) and the 2009-2011 period (blue). Only 
knots with proper motion velocities above 2$\sigma$ (vertical dashed lines) are included.
}
\label{vel41}
\end{figure}

The highest tangential velocities are found for the four knots MHO~2235-E, MHO~2234-A2,
MHO~2234-B1, and MHO~2234-D1+D2, but there is an all over large variation in speed, see 
Sect.~\ref{timevar} for a discussion on time-variable velocities. The median velocities over 
the longest time-scale of 10 years are about 50~km/s, but the scatter is large, from 30 to 
190 km/s. These are based on the measurements of 17 knots only, about 1/3 of all identified 
knots, and about 60\% of the knots we were able to measure. Histograms over the velocity 
distributions are shown in Fig.~\ref{vel41}. For the shorter time-scales we are not able to 
measure the lowest velocities due to the larger error bar, but it is clear that some knots 
show high velocities over short time-scales, while it averages out over longer time-scales. 
This is evidence of time-variable velocities showing that knots speed up and brake down.

\begin{table}
\caption{The 6 different base-lines. The first column gives the base-line over years and the 
second columns shows the sub-indices used, referring to epochs 1, 2, 3, and 4. N denotes the 
number of knots with proper motions larger than 2$\sigma$. The estimated errors in proper motion 
speed ePM$_{\rm ji}$ are converted to errors in the tangential velocities (assuming d=415 pc). }
\label{stat}    
\centering   
{\small                    
\begin{tabular}{lllrrr}        
\hline\hline
Base-line & ij & time  &  N & ePM$_{\rm ji}$ & eV$_{\rm tan}$  \\
          &    & (yr)  &    & (mas/yr)       &    (km/s)        \\ 
\hline  
\\                      
2003-2009 & 12 & 6.040 & 21 &      10        & 20  \\
2009-2011 & 23 & 2.171 & 14 &      28        & 55  \\
2011-2013 & 34 & 1.785 & 10 &      34        & 67  \\
2009-2013 & 24 & 3.956 & 11 &      15        & 30  \\
2003-2011 & 13 & 8.211 & 19 &       7        & 14  \\
2003-2013 & 14 & 9.996 & 17 &       6        & 12  \\
\hline                        
\end{tabular}
}
\end{table}


\subsection{H$_2$ v=1-0 S(1) line fluxes}
\label{var}

For flux measurements, the images are first ``seeing matched'' using the IRAF task {\it gauss} 
which convolves the images with a Gaussian function. All images were matched against the 
``worst seeing'' image using $\sim$ 50 stars. Secondly, ``flux matching'' with respect to ep4 
(2013) was performed using the stellar fluxes. The flux matching error is around 6\%, i.e. this 
is how much the stellar fluxes scatter. We note that many of these stars are Young Stellar 
Objects (YSOs) currently forming in the region \citep{dju06,har06,har07}, and YSOs are known to be 
variable. While a few highly variable stars were discarded before flux matching, low level 
variability in the stellar sample still might have artificially increased this scatter. 
%

The individual knot fluxes were measured in their apertures using the IRAF task {\it phot}. 
The aperture radii varied from 5 pixels ($1.17''$) for the smallest and faintest to 30 
pixels ($7''$) for the largest and most diffuse knot, but stayed mostly between 5 and 9 
pixels. For elongated knots we used large radii and checked manually that nearby knots did 
not contaminate the aperture. For each knot the aperture radius was fixed for all epochs, 
only its position was redefined. The sky annuli were selected depending on the crowding to 
minimize contamination. 

%

\begin{figure}
\centering
\includegraphics[angle=270,width=9cm]{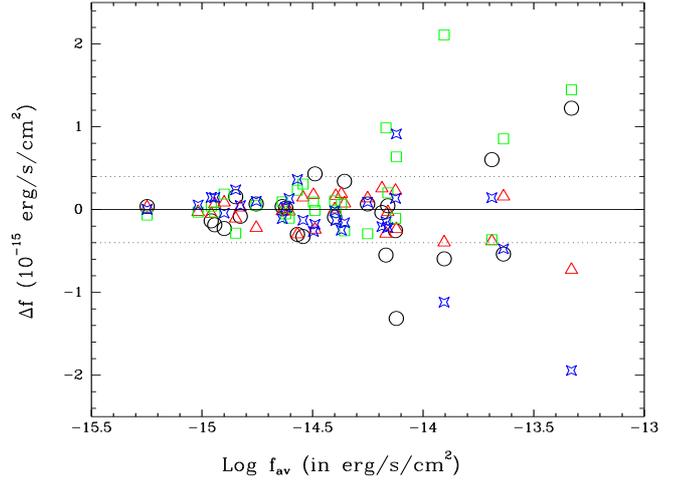}
\caption{Flux variability detected in 13 of the 29 knots. The deviations from 
average flux is plotted versus the logarithm of the average flux for each of the 
knots. Epoch 1 (black circles), epoch 2 (red triangles), epoch 3 (green squares) 
and epoch 4 (blue stars). 
Variability is detected whenever the maximum deviation between epochs is larger
than 3~$\sigma$, where $\sigma$ is the standard deviation of the sky background 
in the epoch 4 image. The dotted line shows 3~$\sigma$ from average. }
\label{fluxvar}
\end{figure}

The flux for the same knot is found to be overall fairly constant in all the four epochs, 
see Table~\ref{fluxes}. To search for variability we calculate the maximum flux variation, 
i.e. the difference between the lowest and highest flux measured for the same knot. A knot 
is defined to have flux variability if the maximum flux deviation is larger than three 
times the standard deviation of the sky background $\sigma_{\rm sky}$ in the deep image of 
ep4, which is $1.5 \times 10^{-16}$~erg~s$^{-1}$~cm$^{-2}$. As shown in Fig.~\ref{fluxvar} we 
find flux variability for 13 out of 29 knots. There is no apparent general correlation 
between flux variability and proper motion. 

\begin{table}
\caption{The H$_2$ v=1-0 S(1) fluxes of the knots for each of the four epochs. This is the 
line emission measured through an aperture with radius R$_{ap}$ (in arcseconds) centred on 
each knot. MHO~2234-A1+A2 comprises both knots although A1 is only clearly emerging in ep4. 
The fluxes f$_1$, f$_2$, f$_3$, and f$_4$ are given in units of 10$^{-15}$~erg s$^{-1}$ cm$^{-2}$. 
The last column shows whether we find the knot flux to be variable or not.}
\label{fluxes}    
\centering  
\input{fluxes2.tex}
\end{table}

\subsection{Radial velocities}
\label{rv}

Since the H$_2$ emission is spatially resolved and the emission lines have narrow essentially 
Gaussian line profiles, we can use profile fitting to obtain a radial velocity resolution much 
higher than the intrinsic resolution of the observations of 140 km s$^{-1}$. For each knot we measure 
the radial velocities from the available H$_2$ lines using the IRAF tasks {\it rvidlines} and 
{\it rvcorrect} to obtain the local-standard-of-rest radial velocities ($v_{LSR}$). The accuracy 
in the wavelength calibration of our spectra is 0.3~\AA \ or 4~km~s$^{-1}$ (see 
Sect.~\ref{specobs}), and since the calibration is done on sky lines in the same spectrum we do 
not expect any shifts. The reference wavelengths for the H$_2$ lines are taken from the vacuum 
wavenumbers provided by \citet{bla87}. The number of available lines vary from 1 to 6, depending 
on the brightness of the knots. 

The measured radial velocities are very similar for all lines in the same knot. The adopted 
error listed in Table~\ref{rvtab} is taken to be the rms scatter in the radial velocity measurements 
line-by-line rather than the mean error. The Serpens NH3 NE cloud core itself has a mean radial 
velocity $v_{\rm LSR} = $ 8.5~km~s$^{-1}$ according to \citet{tob11}, whose measurement is centered 
on the Class\,~0 source MMS3, and 7.986~km~s$^{-1}$ according to \citet{lev13}, who refer to this 
cloud core as Do279P6. We apply 8~km~s$^{-1}$ when correcting for the systemic velocity. We assume 
the protostar that gives rise to the jet follows the cloud motion and thus correct for it to find 
the knot radial velocity with respect to the driving source. The radial velocities given in 
Table~\ref{h2lines}, with their uncertainties and the number of lines used in the measurement, are 
thus the LSR velocities corrected for the systemic velocity of the cloud. 

The negative radial velocities in the southern MHO~2234 jet and the positive radial velocity for 
one measurable knot in the northern, gives the orientation of the jet along the line of sight. 
The small number of northern knots detected is likely due to the fact that the northern jet moves 
away from us and into the cloud. The only northern knot that has measurable radial velocity is 
MHO~2234-J, a bow shock shaped feature outside our H$_2$ imaging field, for which we have no 
proper motion measurements. The knot MHO~2234-A1, for which we have no proper motion because
it was too faint in all epochs except for epoch~4, was bright by the time the spectrum was taken 
was and included in the slit, thus its v$_{\rm rad}$ was obtained together with the one of knot A2. 
These two knots have the largest radial velocity measured.
The radial velocity of MHO~2235-B is very small, but positive, thus the spatial velocity of this 
knot is towards SE and away from us.

\begin{table}
\caption{Radial velocities v$_{\rm rad}$ obtained from the available emission lines, corrected 
for the systemic velocity of the cloud core of $v_{LSR}$ = 8 km s$^{-1}$, and the number of lines 
used to measure v$_{\rm rad}$. The tangential velocities v$_{\rm tan}$ are obtained from the proper 
motion measurements over 10 years and a distance assumption of d = 415 pc. The space velocities 
v$_{\rm space}$ are obtained by combining v$_{\rm rad}$ and v$_{\rm tan}$. }
\label{rvtab}    
\centering   
\begin{tabular}{lrcrr}        
\hline              
Object & v$_{\rm rad}$     & N$_{\rm lines}$ & v$_{\rm tan}$ & v$_{\rm space}$ \\
       & (km s$^{-1}$)     &                 & (km s$^{-1}$) & (km s$^{-1}$)   \\
\hline\hline                       
MHO2234-A1+A2$^{*}$  & -32 $\pm$ 7 & 4               & 121 $\pm$ 12  &  125 $\pm$ 14   \\
MHO2234-D1+D2  &  0 $\pm$ 4  & 4               & 110 $\pm$ 12  &  110 $\pm$ 13   \\
MHO2234-H1    & -12 $\pm$ 8 & 4               &  71 $\pm$ 12  &   72 $\pm$ 14   \\
MHO2234-J     & 16 $\pm$ 1  & 2               &     n.a.      &      n.a.       \\
\hline                         
MHO2235-B     & 5 $\pm$ 3   & 5               &  26 $\pm$ 12  &   26 $\pm$ 12   \\
\hline
\end{tabular}
$^{*}$) Both knots are included in the slit, but v$_{\rm space}$ is valid only for A2.

\end{table}

\subsection{Identification of flows and their driving sources}
\label{flows}

Combining the tangential velocities derived from the proper motions with the radial velocities, 
available for a subset of the knots, we obtain the space velocities. We use the space 
velocities, the morphology, and the locations of known protostars listed in Table~\ref{ysos} to 
evaluate the best candidate driving source for each jet, and also to distinguish between separate 
jet flows and determine which knots belong to which flow. We find at least three separate jet 
flows, partly overlapping, in the field of view studied.

It should be noted that there are several emission line features that can not easily be attributed 
to any of these flows, either because they have no velocity information and/or their location and 
morphology is difficult to interpret. This concerns in particular the features named Q, R, S, and T 
in Fig.\ref{h2image}.

\begin{table}
\caption{Driving source candidates. Values of L$_{\rm bol}$ and T$_{\rm bol}$ and estimates of disk 
masses, as well as positions from CARMA 230 GHz observations are all from \citet{eno09,eno11}. }
\label{ysos}    
\centering   
\begin{tabular}{lrrcrrr}        
\hline              
Ser-emb-   &  $\alpha_{2000}$ & $\delta_{2000}$ & YSO & L$_{\rm bol}$ & T$_{\rm bol}$ & M$_{\rm disk}$ \\
           &   (h)   &  (deg)      & & (L$_{\odot}$) & (K) & (M$_{\odot}$) \\
\hline\hline                       
1 (MMS3)   & 18 29 09.09 & 00 31 30.9  & 0 & 4.1 & 39 & 0.28 \\
11 E       & 18 29 06.75 & 00 30 34.4  & I  & - & - & -   \\
11 W       & 18 29 06.61 & 00 30 34.0  & I  & 4.8 &  77 & 0.15  \\
17         & 18 29 06.20 & 00 30 43.1  & I  & 3.8 & 120 & 0.15  \\
\hline
\end{tabular}
\end{table}

\subsubsection{Flow~1 and MMS3 (Ser-emb-1)}

What we call Flow~1 essentially coincides with the MHO~2234 flow as depicted in Fig.~\ref{h2image}. 
In \citet{dju06} we suggested the driving source to be the Class\,0 source MMS3, also called 
Ser-emb-1 by \citet{eno09,eno11}, who found it to be a single compact source and an early Class\,0, 
probably the least evolved of all protostars in Serpens. From the present study it is confirmed 
that the jet has a bipolar structure with a southern flow (Fig.~\ref{mho2234s}) coming towards 
us and a northern (Fig.~\ref{mho2234n}) going away from us. From the spatial velocities of the 
two knots A2 and H1 in the southern arm we estimate an inclination angle of 10-15 degrees with 
respect to the plane of the sky. This is in agreement with the VLA velocity maps of 
\citet{tob11}, although their axis of the outflow in the plane of the sky is directly 
North-South, while both the morphology of the jet, as seen in near and mid IR images, as well 
as the proper motions of the H$_2$ knots, indicate a sky position angle of about 10 degrees. 

In the southern arm we find a spectacular morphology of at least three main structures, each of 
which consist of curvy whisps of knots that look like fragmented bow-shocks, headed by the B knots 
at $\approx$ 9'' from the driving source, the F knots at $\approx$ 35'' and the H knots at $\approx$ 
60''. We speculate that these multiple bow shocks are the result of several bursts of outflow 
along the jet axis in the recent past, produced by a time-variable jet ejection mechanism. We do not 
see a strong bi-polar symmetry though, probably because the northern arm moves slightly into 
or behind the cloud and is detected at 2 microns only in two locations; 1) a bubbly and 
faint multi-knot feature around the I1-2 knots, at ($\approx$ 37'') from the exciting source, and 2) 
the clear bow shock shaped feature named J, which is at $\approx$ 120'' from the driving source. 
In the 4.5$\mu$m IRAC image, however, there is extended emission right to the north of the
exciting source, most likely the northern component of the bipolar flow moving into the cloud, 
being very reddened (cf. Fig.~\ref{imshift}).

From the velocity information we find it likely that the southernmost part of the MHO~2234 jet, 
containing the H1-H8 knots, is indeed part of the north-south going Flow~1 (MHO~2234), and not part 
of the MHO~2233 structure. Knot H8 apparently moves in the reverse direction, though.

The space velocity of the knots decreases with distance from the originating source, the faster 
knot being MHO~2234-A2 at a space velocity of 125 km s$^{-1}$. Its close neighbour A1 is very 
faint in ep1, and becomes gradually brighter to allow its position to be measured only in epoch 4, 
but both A1 and A2 are included in the flux measurement aperture and in the slit spectroscopy. 
As shown in Sect.~\ref{dynage} where we calculate the dynamical ages, the jet from the least 
evolved protostar shows higher knot velocities than the jet from the more evolved protostars. 
This is in line with \citet{bont96} who found roughly an order of magnitude larger 
mechanical-to-bolometric luminosity ratio for outflows from Class\,0s compared to Class\,Is. 

\begin{figure}
\centering
\includegraphics[width=9.2cm]{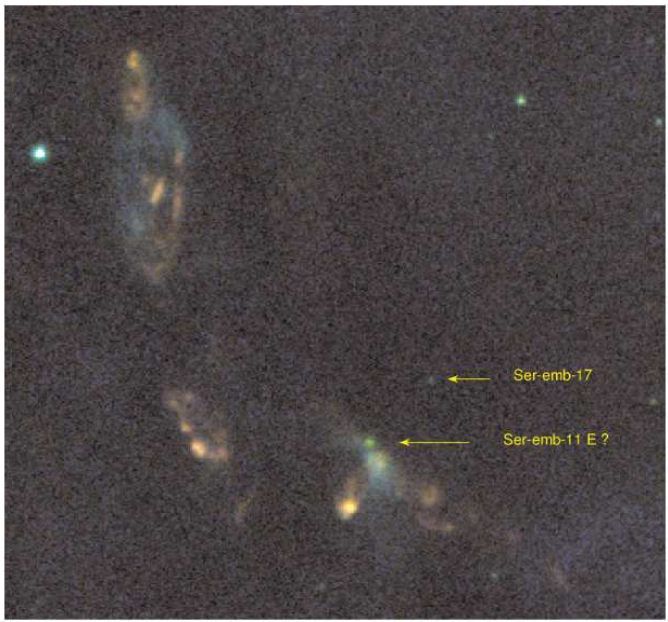}
\caption{Central part (FOV $\approx 90''$ ) of the jets colour coded $\mathrm{H}_2$ line filter 
(red), K-band filter (green) and H-band (blue). North up, East left. }
\label{ne_rgb_h2kh}
\end{figure}

\subsubsection{Flow~2 and Ser-emb-11 binary, E and W}

What we call Flow~2 is the longest structure in our 4' field of view, a NE-SW oriented flow 
composed of parts of both MHO~2233 (clearly N and P, and probably O) and MHO~2235 (clearly C1, D, 
E, F, and probably C2, I1, I2, I3). The previously known HH object HH~476, which practically 
coincides with our group of M knots, is located at an extension of this line, but the proper motion 
of knot MHO~2235-M4 has a consistent reverse movement, and it is unclear if this group of 
knots belong to Flow~2. We also note that the knots MHO~2235-G, H, J and K could be related to 
Ser-emb-34, in which case it would be a strongly deflected flow, with no detectable counterjet, but 
this speculation, however, awaits proper motion measurements.
As seen in Fig.~\ref{irac}, and also as noted in the MHO Catalogue \citep{dav10}, MHO~2235 may be 
extending further towards the SW far outside the NOTCam field. Whether this is an extension 
of Flow~2 or an independent flow, is not clear, as we have no velocity information, but no 
known protostars are found close to these features. The morphology suggests a possible connection, 
but if both structures to the SW are part of the same flow, this indicates a strong deflection 
of the jet axis. Thus, Flow~2 could be a parsec scale flow, with parts of it completely void of 
emitting knots indicating periods of quiescence in the jet ejection. 

Considering the field we have monitored, the knots flowing towards the southwest have very clear 
proper motions and include MHO~2235-E moving at 190 km s${-1}$, the fastest knot in this study. 
The flow towards northeast is more difficult to measure, and only MHO~2233-N has a clear proper 
motion towards the northeast. From the morphology and tangential velocities, we suggest that this 
flow originates in Ser-emb-11, or in either of the two components of this $2.14''$ separation 
Class\,I binary. 

Inspecting our supplementary deep JHK$_S$ imaging taken simultaneously with H$_2$ line imaging and 
a narrow-band K continuum filter, we find evidence that our knot MHO~2235-A2, which almost 
coincides with the protostar Ser-emb-11E, indeed has emission both in the red continuum filter, 
as well as in the H$_2$ line filter. This source is the eastern and the fainter of the two 
Class\,~Is that form the 890 AU separation protostar binary Ser-emb-11, which is found by 
\citet{eno11} to have very extended envelopes and compact disk components. In our images 
MHO~2235-A2 has a roundish shape, but its psf in the best quality images, where stars have a FWHM 
of $0.5-0.6''$, is substantially broader.  The positions disagree by 0.4''. The binary companion 
Ser-emb-11W, which is the brighter of the two in submm wavelengths, is not detected in any near-IR 
filter. We find no measurable proper motion from MHO~2235-A2. On the other hand we detect a gradual 
brightening in flux over the 10 years of monitoring in the H$_2$ line filter, see Tables~\ref{pmtab} 
and \ref{fluxes}. Thus, what we refer to as knot MHO~2235-A2 may be a mix of continuum emission from 
the YSO itself (Ser-emb-11E) and some extended line emission. 

As shown in Fig.~\ref{ne_rgb_h2kh} there is a cone of faint emission extending southwards from 
Ser-emb-11E (at about 5 o'clock) in the K continuum filter and in the H-band. This emission is barely 
noticable in the J-band, probably due to high extinction. Such cones of scattered light are usually 
interpreted as cavities in the envelope carved out by a precessing jet. The part of Flow~2 that we 
have monitored is a long chain of knots with a slightly S-shaped morphology. This point symmetry is 
usually interpreted as due to a precessing jet axis. As shown by \citet{fen98} this is most likely 
caused by orbital motion of the jet source in a binary (or multiple) system. 

\begin{figure}
\centering
\includegraphics[width=9.2cm]{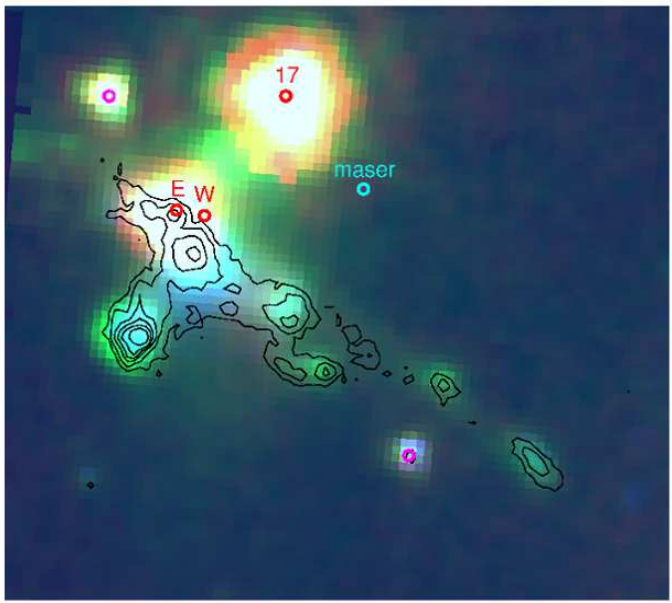}
\caption{The central $\approx 45''$ zoom-in of IRAC/Spitzer image shown in Fig.~\ref{irac}. 
Protostars are marked and labelled. The black contours refer to the 2.1218 $\mu$m H$_2$ 
line emission.}
\label{iraczoom}
\end{figure}

\subsubsection{Flow~3 and Ser-emb-17}

Proper motion and radial velocity data show that MHO~2235-B, the brightest of all knots
in the images, is moving towards the South-East and slightly away from us. From the radial 
velocity component we determine the inclination angle with respect to the plane of the sky 
to be 11$^{\circ}$. Knot MHO~2235-A1 has a similar, within the errors, proper motion vector. 
These two knots seem to form a separate flow approximately orthogonal to the above described 
Flow~2. The Class\,I binary Ser-emb-11 is situated approximately where the two flows cross 
each other in the plane of the sky. Thus, it is likely that the south-east going Flow~3 is 
either driven by one of the components of the Class\,I binary Ser-emb-11 or by Ser-emb-17, 
the proper motion vectors being consistent with both. The Class\,I source Ser-emb-17 
\citep{eno09,eno11} is located only 12'' to the North-West of the binary Ser-emb-11 and is 
detected as a faint point-source in the K-continuum and the Ks broad band filters only, see 
Fig.~\ref{ne_rgb_h2kh}. Careful inspection of the IRAC image in Fig.~\ref{iraczoom} reveals 
a {\em bridge} of H$_2$ emission (green colour gives an emphasis to the 4.5 micron band) 
between Ser-emb-17 and Ser-emb-11, suggesting a flow in the direction towards MHO~2235-B. 
The non-detection of this extended emission at two microns is probably due to very high
extinction. Ser-em-17 is located in the densest core in the region according to the N$_2$H$^+$ 
map of \citet{tob11}. Extinction may also explain why we detect no bipolar lobe towards the 
north-west. Based on this and the morphology and velocity of MHO~2235-B, we propose Ser-emb-17 
to be the driving source of the south-east flowing jet. 
 
Based on jet lengths only, one would wrongly interpret Flow~3 as the youngest jet in the region. 
Including the velocity information, however, we find that the furthest away knots in the longer 
Flow~1 from the Class\,0 source has approximately the same dynamical age, see Sect~\ref{dynage}. 
A recent study by \citet{gia13} finds a clear correlation of jet length with stellar age, but 
there were no Class\,0 YSOs in their sample of driving sources.

\subsection{Time variable velocities}
\label{timevar}

It is evident from the multi-epoch proper motion results presented in Table~\ref{pmtab}
that the proper motions are not constant over time. Translated to tangential velocities
using 415 pc for the distance, the most typical velocity, averaged over 10 years, is 
around 50 km/s, and the highest velocity is 190 km/s as shown in Fig.~\ref{vel41}, but 
short-term proper motions over $\sim$ 2 years reveal ``transient'' velocities of up 
to 300 km/s, for instance MHO~2234-B1 from 2009 to 2011 and MHO~2235-E from 2011 to 2013. 
These are knots that have speeded up but move in the same direction as for the other 
periods. Thus, the H$_2$ knots accelerate and decelerate.

The observable knots are shocks, probably produced by supersonic velocity jumps, for which 
a natural origin is believed to be a time variable jet ejection speed. Episodically pulsed 
jets, due to non-steady accretion in the form of bursts, was suggested as an explanation of 
knotty jet structures by \citet{rei89}, and modeled by \citep{rag90}, and strong evidence for 
this mechanism was found in the highly symmetric knots in the HH~212 bipolar jet \citep{zin98}. 
The morphology and the knot spacings indirectly point towards time-variable velocities. Direct 
measurements of time-variability in jet velocities, however, are scarce. This has been detected 
from multi epoch studies of a Class\,II jet in HH~30 \citep{hart07}, a Class\,I binary jet in 
the L1551 \citep{bon08}, and Class\,II and Class\,I jets in the Chamaeleon II star formation 
region \citep{car09}. It is suggested that Class\,0 jets have a similar variability behaviour, 
based on observed knot spacings, but to our knowledge, the present study in Serpens is the first 
that finds time-variable velocities in a Class\,0 jet. 

Ejecta at different velocities propagating in a strongly inhomogeneous medium give rise to rather 
complex interactions and possibly deflections. The jets in Serpens NH3 (Serpens Cluster B) are 
partly overlapping, and it is not simple to interpret which knot belongs to which flow, 
considering reverse motion and deflections.

It is believed that faster ejecta may overtake slower ones, and potentially change their velocity 
vectors. Models of randomly pulsed jets by \citet{bon10} predict: distinct individual knot 
velocities, single fast-moving knots, reverse shocks, as well as oblique shocks, the latter being 
produced by reflection at the cocoon surrounding the jet. Our velocity data for the knots in these 
jets suggest the presence of several of these features. The distinct individual knot velocities are 
clearly demonstrated by our proper motion results. An example of a single fast-moving knot is 
MHO~2235-E. Possible examples of reversing knots are MHO~2235-M4 and MHO~2234-H8.  

Whether a variable jet ejection mechanism is needed to produce the observed time variability 
in the knot velocities is not enirely clear, and \citet{yir09,yir12} suggest that shocks can 
be produced by clumps in the jet that collide and merge. If the shocks are due to merging or 
interacting knots, one would expect variability in the fluxes, as well. The fast moving 
elongated knot MHO~2235-E experiences both velocity and flux variability, but in general we 
find no clear correlation between flux variability and velocity variability, nor between flux 
variability and speed.


\section{Physical conditions}
\label{physical}

\begin{figure}
\centering
\includegraphics[angle=270,width=9cm]{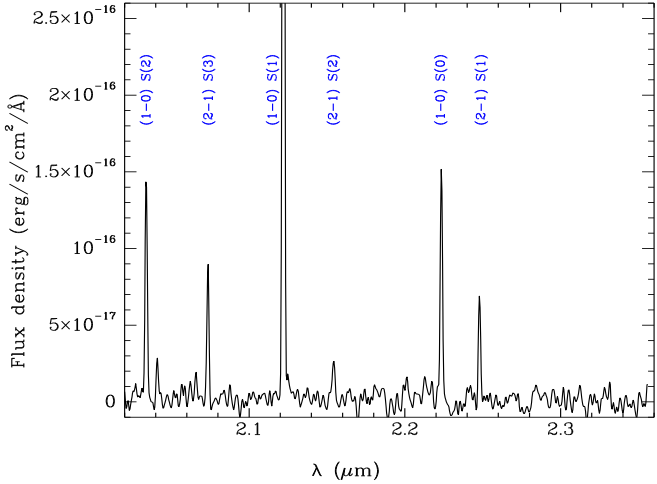}
\caption{NOTCam K-band (R=2100) spectrum (slightly smoothed) of the brightest knot MHO~2235-B 
with the H$_2$ lines marked. Flux density in units of erg~s$^{-1}$~cm$^{-2}\AA^{-1}$. }
\label{kspec}
\end{figure}

The knots for which we have K-band spectra are listed with their line fluxes or flux ratios 
in Table~\ref{h2lines}. The slit positions are shown overlaid on images in 
Figs.~\ref{mho2235cen}, \ref{mho2234s}, \ref{mho2234n} and \ref{mho2233east}. All the knots 
have pure emission line spectra with narrow line profiles, the gaussian widths being of the 
order of the instrument profile, around 10 \AA \ or 140 km s$^{-1}$. Up to six or seven H$_2$ 
lines are typically detected in the spectra. There are no other lines, nor continuum emission. 
Figure~\ref{kspec} shows knot MHO~2235-B as an example. For all knots the brightest line is 
the v=1-0~S(1) line at 2.1218 $\mu$m which contributes on the average 59 $\pm$ 17 \% of the 
total emission in the $K$ band.

\begin{table*}
\caption{Measured H$_2$ line fluxes and line ratios. The flux density (in erg~s$^{-1}$~cm$^{-2}$) 
is given for the v=1-0 S(1) line at 2.1218 $\mu$m, and flux ratios with respect to this line 
are given for the other lines. The third last column gives the estimated excitation temperatures 
T$_{ex}$, the second last gives the estimated extinction (A$_V$), and the last column gives the 
calculated ortho to para ratio $o/p$ for the v=1 vibrational level.}
\label{h2lines}    
\centering   
\begin{tabular}{lrcccccccc}        
\hline              
Object& 1-0 S(1)           & 1-0 S(2)      & 2-1 S(3)       & 2-1 S(2)      & 1-0 S(0)      & 2-1 S(1)      &T$_{ex}$        & A$_V$ & o/p \\
      & 2.1218 $\mu$m      & 2.0338 $\mu$m & 2.0735 $\mu$m  & 2.1542 $\mu$m & 2.2233 $\mu$m & 2.2477 $\mu$m &                &       &     \\
  & erg s$^{-1}$ cm$^{-2}$ & ratio         & ratio          & ratio         & ratio         & ratio         & K              & mag   &     \\
\hline  
\hline                   
MHO~2235-B   & 1.4 10$^{-14}$ &0.32$\pm$0.02 & 0.10$\pm$0.03  &  0.06$\pm$0.07& 0.26$\pm$0.02 & 0.13$\pm$0.04 & 2240 $\pm$ 160 & 23    & 2.9 \\
\hline
MHO~2234-A1+A2& 4.4 10$^{-15}$ &0.23$\pm$0.04 & 0.04$\pm$0.20  &  -            & 0.30$\pm$0.05 & 0.14$\pm$0.07 & 1800 $\pm$ 310 & 58    & 3.2 \\
MHO~2234-D1+D2& 5.9 10$^{-15}$ &0.27$\pm$0.04 & 0.12$\pm$0.07  &  -            & 0.19$\pm$0.06 & 0.19$\pm$0.05 & 2500 $\pm$ 300 & 14    & 3.7 \\
MHO~2234-H1  & 5.6 10$^{-15}$ &0.22$\pm$0.07 & 0.04$\pm$0.09  &  0.04$\pm$0.12& 0.19$\pm$0.06 & 0.06$\pm$0.11 & 1900 $\pm$ 170 & 31    & 4.1 \\
MHO~2234-J   & 1.3 10$^{-15}$  & -          &   -            &  -            &  -            & 0.36$\pm$0.20 &      -         &  -    & -    \\
\hline
MHO~2233-N   & 3.3 10$^{-15}$ &0.35$\pm$0.05 & 0.16$\pm$0.10  & 0.07$\pm$0.19 & 0.23$\pm$0.10 & 0.20$\pm$0.11 & 2620 $\pm$ 160 & 7    & 3.0 \\
\hline                        
\end{tabular}
\end{table*}

\subsection{Extinction}
\label{ext}

There is substantial cloud extinction in the Serpens~NH3 core where these jets are located, and 
the region is practically opaque in the optical. Even in the K-band there is a pronounced decrease 
in the number of background stars over the region where the jets are located. A lower estimate of 
A$_V~\sim$~20~mag was deduced from the $J-H/H-K$ colour diagram of stars in the surrounding region 
\citep{dju06}, which agrees well with the extinction map of \citet{bont10}, but these methods 
``saturate'' for the densest regions. There is no emission from the knots in the J and H
bands. Deep H band imaging shows some faint extended emission between some the knots, which we 
interpret as due to scattered light. 

Assuming thermalized gas at a single temperature, we can use the pair of lines originating from 
the same upper level that are most separated in wavelength, $\lambda\lambda$ 
2.0338 and 2.2233 $\mu$m, in order to determine the colour excess between these two wavelengths. 
We then apply the extinction law A$_{\lambda} \propto \lambda^{-1.7}$ to deduce the extinction in 
terms of visual magnitudes. We have in this manner estimated the extinction to the knots 
(Table~\ref{h2lines}) to be in the range $7 < A_V < 58$ magnitudes. These values are obtained 
over a small wavelength range, however, and their uncertainty may be large. The highest 
estimate ($A_V = 58$ mag) was found for the knot closest to the Class\,0 exciting source. Such 
high values are rarely found in the literature, and above all difficult to measure. 
\citet{dav11} compared various techniques to measure the extinction towards molecular hydrogen 
emission line regions around protostars using integral field spectroscopy and found values reaching 
up to $A_V \sim 80$ mag close to the protostellar jet sources. In another study of knots from 
protostellar jets by \citet{car15} an extinction of $A_V = 50$ mag is reported for one of the 
knots. Knot emission seen through 50 visual magnitudes of extinction ($\sim$ 5 magnitudes at 
two microns) is weakened by a factor of 100 and must be intrinsically bright.

With this high extinction, one could expect that some knots would be extinguished at 2.1 $\mu$m
microns, but become observable at 4.5 $\mu$m where the extinction is lower, for instance 
\citet{fla07} find that A$_{[\rm 4.5]}$ = 0.53 A$_{K}$.
When comparing the areas covered by $H_2$ line imaging with the IRAC images, we conclude that 
practically all knots found in the mid-IR images are also visible in the near-IR 2.122 $\mu$m 
images, except for two important cases where we believe the extinction is very high:
1) just to the north of the Class\,0 exciting source and 2) between MHO2235-B and its 
presumably driving source Ser-emb-17. Otherwise, the practically one-to-one 
correspondence between IRAC and ground-based 2.122 $\mu$m images, is also in line with what 
\citet{gia13} finds for the Vela-D molecular cloud, and this suggests that even for relatively 
high extinction regions, such as in Serpens NH3, near-IR $H_2$ line mapping is able to pick up 
most of these emission line objects.

\subsection{Excitation mechanism}

The excitation mechanism of H$_2$ in molecular jets from young stars is mainly due to collisions 
with other H$_2$ molecules or with H atoms, where typically the lower vibrational levels of the 
molecule in its ground state get populated. If UV photons are present, however, also the higher 
vibrational levels can be populated through radiation pumping of $H_2$ into a higher electronic 
state, followed by a subsequent decay that populates all the vibrational levels of the ground 
state. A frequently used rule of thumb is that when the ratio $v=2-1~S(1)/v=1-0~S(1)$ 
(2.1477/2.1218) is $\la$ 1/10 and there is an absence of v~$\ge$~3 lines, the excitation mechanism 
of $H_2$ is from collisions in a hot gas \citep{wol91}, i.e. due to shocks. 

\begin{figure}
\centering
\includegraphics[angle=270,width=9cm]{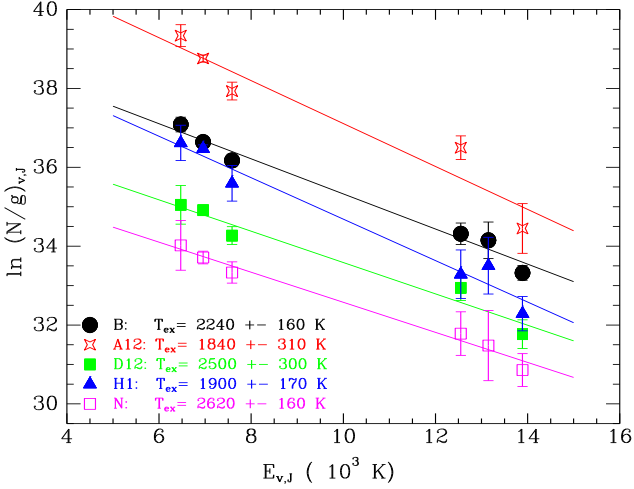}
\caption{Excitation diagram for the knots listed in Table~\ref{h2lines}. The entries for each 
knot are marked with different symbols and colours and are calculated from the de-reddened line 
intensities in units of erg s$^{-1}$ cm$^{-2}$ sr$^{-1}$ according to the relation given in the 
text. Linear fits are made to the points for each knot. 
Column densities $N(v,J)$ are in the units of cm$^{-2}$.
}
\label{boltzmann}
\end{figure}

In order to estimate the excitation temperature and the column densities of H$_2$ 
in the knots, we construct the Boltzmann diagrams as described in \citet{nis08}, where we
depart from the Boltzmann distribution that describes a gas in thermal equilibrium

\begin{equation}
\frac{N(v,J)}{N(H_2)} = \frac{g_{v,J}}{Z(T)} e^{-E(v,J)/kT_{\rm ex}}
\end{equation}

to get that the natural logarithm of $N(v,J)/g_{v,J}$, where $g_{v,J}$ are the statistical weights 
from the selection rules of the transitions, equals $- E(v,J)/kT_{ex}$. Thus, when plotting 
$\ln ((N(v,J)/g_{v,J}$) versus $E(v,J)$, a thermalized gas will give a linear relation, where the 
slope is the inverse of the excitation temperature $T_{ex}$. $Z(T)$ is the partition function,
and the ro-vibrational energies $E(v,J)$ are taken from \citet{dab84}. 
To calculate the relative column densities $N(v,J)$ for a given vibrational and rotational level 
$(v,J)$, we use the relation $N(v,J) = 4\pi I(v,J)/(h\nu g(v,J) A(v,J)$, where I(v,J) 
are the measured line intensities obtained from the line fluxes given in Table~\ref{h2lines}, 
corrected for extinction using the derived extinction for each knot, and given in units of 
erg s$^{-1}$ cm$^{-2}$ sr$^{-1}$ using that the spectra cover an area of 0.6$\arcsec \times$ 
1.9$\arcsec$ across each source. The Einstein coefficients $A_{v,J}$ are from \citet{tur77}. 
We fitted a line through the points for each knot, as shown in Fig.~\ref{boltzmann} and 
the resulting $T_{ex}$ values and errors are listed in Table~\ref{h2lines}. Since both the $v=1$ 
and $v=2$ vibration levels can be fitted with the same slope, this suggests that the gas is in 
LTE and the excitation mechanism is through collisions. The average excitation temperature 
T$_{ex}$ is 2200 $\pm$ 350 K, which is within the typical range of gas temperatures in post-shock 
regions of protostellar jets (2000-4000 K). 

\subsection{Ortho to para ratio}

When H$_2$ molecules are formed, their nuclear spins can either be aligned (ortho) or opposed 
(para). The ortho to para ratio of H$_2$, referring to these two possible spin states, has 
been estimated for the vibrationally excited state v=1 using the line ratios 
v=1-0~S(1)/v=1-0~S(0) and v=1-0~S(1)/v=1-0~S(2) following the method of \citet{smi97}. For each 
knot we applied the individual extinction estimates as found above, though extinction has 
marginal effect on the results. Our values for the ortho to para ratios are listed in 
Table~\ref{h2lines} and vary from 2.9 to 4.1 with an average value of 3.3, which is consistent 
with the expected ratio of 3:1 based on the statistical weights. This very ratio is found in 
many regions of Herbig-Haro flows. The interpretation could be either 1) that the molecules 
formed on grains at high temperature according to the statistical weights, and there has been no 
conversion since, or 2) that collisions with H atoms in the shocked gas has thermalized the 
level populations and produced this ratio. According to \citet{smi97} the latter is the more 
likely explanation.

\subsection{H$_2$ column densities}
\label{coldens}

For the five knots that have spectra taken, we derive the column densities of H$_2$ from
the already constructed Boltzmann diagram in Fig.~\ref{boltzmann}. We obtained the values of 
ln(N$_{H_2}$/Z(T)) by reading off the intercept of each slope at zero E$_{v,J}$. The partition 
function of molecular hydrogen was calculated as $Z(T) = 0.024 T_{\rm ex}/(1-e^{-6000./T_{\rm ex}})$ 
as given in \citet{smimac97}.

For the remaining knots we have no spectra and are limited to flux measurements obtained 
from H$_2$ narrowband imaging in the $\lambda$ = 2.1218 $\mu$m line. We follow the method of 
\citet{dav01} consisting in using this line intensity together with an assumption on the 
excitation temperature and the extinction for each knot. We use the average of the four flux 
measurements given in Table~\ref{fluxes}.
The excitation temperature varies from 1800 to 2620 K among the five knots studied with 
spectra, and we apply the average value of T$_{\rm ex} = 2200$ K for all other knots. 
The extinction apparently varies strongly on small spatial scales and will be the main 
source of uncertainty using this method. We have estimated the extinction for each knot from 
a rough spatial interpolation relative to positions for which A$_V$ is already measured with 
spectroscopy as described in Sect.~\ref{ext}. This is the best we can do with the available
data, and we caution that an error of 10 magnitudes in the extinction estimate would lead to 
an uncertainty factor of 2.8 in the dereddened line intensities and subsequently in the column 
densities and the luminosities.

The surface area for each knot was estimated individually due to their different morphology, 
and for this we use the deep image from epoch 4. After having subtracted a smoothed version, 
using a 13 $\times$ 13 pixel median of the image, for each knot we count those pixels which 
have fluxes above 3.4 $\times$ 10$^{-16}$ erg~$s^{-1}$~cm$^{-2}$~arcsec$^{-2}$, which is 2.3 times 
the standard deviation of the background. The resulting surface areas for the 29 knots are in 
the range 1.3 to 10.9 square arc seconds, translating to the range from 5 $\times 10^{31}$ to 
4 $\times 10^{32}$ cm$^{2}$ at 415 pc distance, with a median value at 1.3 $\times 10^{32}$ cm$^{2}$. 

We now have dereddened intensities per steradian in the 
v=1-0 S(1) line, which is used to calculate the column densities as described in \citet{dav01}. 
These are listed in Table~\ref{mass-dens} in units of H$_2$ molecules per cm$^{-2}$. The knots 
are grouped by flows, the first block containing the Flow~1 knots, the second block the Flow~2
knots, and the third block the Flow~3 knots. For the purpose of comparison we list the results 
obtained by both methods for those five knots that have spectra, marking the results from 
spectra with an asterisk. The N$_{H_2}$ obtained from imaging compare reasonably well to the more
accurate estimates from spectroscopy, the column densities are typically within a factor of 2-3,
except for knot MHO~2234-A1+A2 where they differ by a factor of five. Thus, the results for knots
where we have no spectra, should be regarded as approximate and accurate only to within an order 
of magnitude. The H$_2$ column densities listed in Table~\ref{mass-dens} 
compare well with the range of values found in similar studies of knots in protostellar flows. 
\citet{car15} found H$_2$ column densities from $10^{17}$ to 2 $\times 10^{20}$ cm$^{-2}$ for a 
sample of 65 knots in flows around 14 sources.

\begin{table}
\caption{Molecular hydrogen column densities as measured using K-band spectra (5 knots) and 
narrow-band H$_2$ line imaging (29 knots). The masses are obtained from the column densities
N$_{H_2}$ and the surface areas of the knots. The mass-loss rates are estimated using the 
velocity information of the knots and their location relative to the driving source. }
\label{mass-dens}    
\centering   
\input{mass_ordered2.tex}
\end{table}

The number densities of molecular hydrogen, $n_{H_2}$, are estimated from the column
densities by assuming that the knots' dimension in the radial direction is equal to their 
narrowest side seen in the images and measured to be on the average 1.4$\arcsec$. The knots 
typically have an elongated shape, mainly along the flow direction, which is roughly the 
orientation of the slits, as well. As learned from the relatively small radial velocities, 
these jets are flowing practically in the plane of the sky. Thus, applying 1.4$\arcsec$ 
translates to a jet diameter of 580 AU. The derived densities $n_{H_2}$ for each knot 
are in the range 30 - 17000 cm$^{-3}$, with a median value around 400 cm$^{-3}$. 
These densities are very low and indicate either that a small percentage of the jet is 
molecular, or that the jets are lighter than the ambient medium.

\subsection{Dynamical ages}
\label{dynage}

On the assumption that velocities are constant, one can calculate the time it has taken for 
(a part of) a jet to move to its current displacement from the originating source by dividing 
jet length by velocity. This represents a dynamical age for the observed feature. 
In the case that velocities slow down with time, the ages are upper limits. We take the jet 
lengths to be the straight line 
projections from knot to driving source, corrected for line of sight inclinations, but not for 
any possible deflections. We use the average tangential velocities over 10 years coupled with 
the radial velocities (whenever measurements exist) to get an approximation of the current space 
velocities of the various parts of the jets. Where we have no radial velocity information, we 
use the projected length. Table~\ref{dynage} lists our dynamical age estimates for a few features 
in each of the three flows and demonstrates how recent the jet activity in this region is. 

\begin{table}
\caption{Dynamical ages of the knots based on their jet lengths, i.e. separation from the 
candidate driving source, and their current velocities, v, where we adopt the space velocities 
when available, and otherwise the tangential velocities over the longest time-scale available. 
The flow for each knot is given in column 2.}
\label{dynage}    
\centering   
\begin{tabular}{lllrrr}        
\hline              
Knot          &  F  & Driving YSO       & Length &  v     & $\tau_{\rm dyn}$ \\
              &      & candidate         &  (pc)  & (km/s) &     (yrs) \\
\hline\hline                       
MHO~2234-A2   &  1   & MMS3              & 0.009  &   125    &     70    \\
MHO~2234-B1     &  1   &                   & 0.019  &   116    &    160    \\ 
MHO~2234-D1+D2   &  1   &                   & 0.048  &   110    &    420    \\
MHO~2234-H1     &  1   &                   & 0.123  &    72    &   1680    \\
MHO~2234-I1     &  1   &                   & 0.078  &    40    &   1900    \\
\hline                         
MHO~2233-N      &  2   & Ser-emb-11        & 0.174  &    42    &   4050    \\ 
MHO~2235-I2     &  2   &                   & 0.208  &    74    &   2750    \\
MHO~2235-M4     &  2   &                   & 0.264  &    44    &   5860    \\
\hline
MHO~2235-A1     &  3   & Ser-emb-17        & 0.028  &    54    &    507    \\
MHO~2235-B      &  3   &                   & 0.044  &    26    &   1650    \\
\hline
\end{tabular}
\end{table}

\subsection{Mass loss rates}
\label{massloss}

From the column densities, N$_{H_2}$, we can estimate the mass of the H$_2$ gas 
involved in the emission through the relation 

\begin{equation}
M_{H_2} = 2\mu m_{H} N_{H_2} a_{\rm knot},
\end{equation}

where $\mu$ is the mean atomic weight, $m_H$ the proton weight, and $a_{\rm knot}$ the area 
of the emitting region \citep{nis05}. The slit width is substantially narrower than the 
FWHM of the intensity profiles across the knots. We assume that the knots emit uniformly 
over their entire surface and use the surface areas derived for each knot as described in 
the previous section. Thus, we arrive at $H_2$ gas masses in the range from 
$2 \times 10^{-8}$ to 7 $\times 10^{-5}$ M$_{\odot}$ for the knots (see Table~\ref{mass-dens}).
The total mass of H$_2$ gas in each flow is obtained by summing up all the measured knots
and listed in Table~\ref{sum}. These estimates are a lower limit to the total H$_2$ gas 
mass. There are a number of weaker knots for which we do not have reliable flux 
measurements. It is clear, however, that there is an order of magnitude more mass in the 
H$_2$ knots of the Class\,0 jet (Flow~1) compared to the two Class\,I jets.

To estimate the H$_2$ mass-loss rates we use the dynamical age, t$_{\rm dyn}$, derived from 
the spatial velocities and the separation of each knot to its putative driving source 
in Table~\ref{dynage} and find per knot \.{M}$_{\rm out}$(H$_2$) = M$_{H_2}$/t$_{\rm dyn}$.
For the 24 knots where we have no radial velocity, we have used the average tangential 
velocities, which for all jets are the dominating vectors of the space velocities, as these 
are all flowing practically in the plane of the sky. In this way we obtain the mass-flux 
carried by each knot. As shown in Table~\ref{mass-dens} the derived mass loss rates per knot 
are in the range 
$4 \times 10^{-12}$~M$_{\odot}$~yr$^{-1}~<$~\.{M}$_{\rm out}$(H$_2$)~$<~10^{-6}$~M$_{\odot}$~yr$^{-1}$. 
Except for the lowest rates, these values compare well with the mass flow rates estimated 
for knots in HH~52, HH~53 and HH~54 \citep{car09}. They also compare well to the mass loss 
rates of $6 \times 10^{-10}$ to $2 \times 10^{-6}$~M$_{\odot}$~yr$^{-1}$ obtained for 9 embedded 
Class\,I outflow sources by \citet{dav01} although these values are derived in the vicinity 
of the driving sources. Likewise, \citet{car12} find an average \.{M}$_{\rm out}$(H$_2$) of 
3.4$\times 10^{-9}$ M$_{\odot}$ yr$^{-1}$ for five Class\,Is and six very young YSOs. We have
summed up individual knot values to arrive at total values for each flow in Table~\ref{sum}.

The derived mass loss rate is two to three orders of magnitude larger for the Class\,0 
jet (Flow~1) than for the two Class\,I jets. This is a combination of the higher mass in 
the ejecta and the higher knot velocities measured for the Class\,0 jet. The momentum flux 
\.{P} = \.{M}v$_{\rm space}$ for the Class\,0 jet (Flow~1) is 1.3~$\times 10^{-4}$~M$_{\odot}$ 
km s$^{-1}$ yr$^{-1}$, which is almost 3 orders of magnitude larger than the momentum flux in 
the Class\,I jets. We note that the the three driving sources have a similar bolometric 
luminosity $L_{\rm bol} \sim 4 L_{\odot}$ \citep{eno11}. Thus, among three YSOs with a similar 
$L_{\rm bol}$ the Class\,0 jet has significantly larger \.{M}$_{\rm out}$(H$_2$) and 
\.{P}$_{\rm out}$(H$_2$), suggesting that the difference is due to the earlier evolutionary 
stage. This is in line with the results of \citet{bont96} who found a stronger CO outflow 
momentum from Class\,0s than from Class\,Is and attributed it to a decrease in the mass 
accretion rate with increasing age of the star. Studies of a larger sample of 23 Class\,0 
and Class\,Is by \citet{car06}, on the other hand, found no clear distinction between the 
two YSO classes when examining outflow $H_2$ luminosity versus $L_{bol}$. For comparison, we 
estimate the outflow $H_2$ luminosity from the dereddened flux in the 2.12 $\mu$m line, and 
get $L_{2.12}$ values of 0.025, 0.0016 and 0.0034~$L_{\odot}$ for Flow~1, Flow~2 and Flow~3, 
respectively. 
The correction factor to apply to $L_{2.12}$ in order to get $L_{H_2}$ is $\sim$ 12 when 
assuming LTE and an average temperature of 2200 K \citep{car06} and give the values listed
in Table~\ref{sum}. 
 
\begin{table}
\caption{Estimates of total $H_2$ luminosity, mass, mass flow rate and momentum flux in 
each flow, as obtained by summing over the individual knots. L$_{\rm bol}$ of the driving
source are from \citet{eno09,eno11}. }
\label{sum}    
\centering   
\begin{tabular}{lccccc}        
\hline              
Flow  & L$_{\rm bol}$ & L$_{H_2}$     & M$_{H_2}$     & \.{M}$_{\rm out}$ & \.{P}$_{H_2}$ \\
      &(L$_{\odot}$) & (L$_{\odot}$) & (M$_{\odot}$) & (M$_{\odot}$ yr$^{-1}$) & (M$_{\odot}$ km s$^{-1}$ yr$^{-1}$) \\
\hline\hline                       
1    & 4.1         & 0.3  & 8.7 $\times 10^{-5}$ & 1.0 $\times 10^{-6}$   & 1.3 $\times 10^{-4}$  \\
2    & 4.8         & 0.02 & 2.0 $\times 10^{-6}$ & 2.0 $\times 10^{-9}$   & 2.6 $\times 10^{-7}$  \\
3    & 3.8         & 0.04 & 8.6 $\times 10^{-6}$ & 3.3 $\times 10^{-9}$   & 1.9 $\times 10^{-7}$  \\
\hline
\end{tabular}
\end{table}

As discussed in Sect.~\ref{timevar} our multi-epoch proper motions show that the
velocities are time-variable. Thus, all discussions on mass flux here is based on the 
use of an average velocity over a 10 year time-span for each knot. 
Within the Class\,0 flow there is a large variation in the estimated mass flow rates 
of the individual knots. Above all others, MHO~2234-A1+A2, which is the knot nearest 
to the Class\,0 driving source (at a projected distance of $\sim$ 1800~AU), stands out 
as particularly massive and fast-moving. This double-knot alone carries a mass flux of 
almost 1~$\times$ 10$^{-6}$~M$_{\odot}$~yr$^{-1}$ and is the main contributor to the 
relatively high \.{M}$_{\rm out}$ in this flow. It is probably the result of a recent burst 
of mass accretion, as these mechanisms are believed to be closely coupled.

Still, according to observed relations between the bolometric luminosity of 
Class\,0/Class\,I sources and their mass flow rate from CO measurements \citep{she96}, 
the expected \.{M}$_{\rm out}$ for a 4 L$_{\odot}$ Class\,0 source should be around 
$10^{-5}$~M$_{\odot}$~yr$^{-1}$. 
This discrepancy is probably explained by the fact that our estimates concern only that 
part of the gas that is involved in the H$_2$ line emission, that is the warm molecular 
gas component. From these H$_2$ data alone, we have no information about the presence 
and contribution to the mass-flow rate of a colder molecular gas component or an atomic 
gas component. Although our deep imaging finds no emission from the knots in the H-band 
where strong [FeII] lines are located, this could be due to high extinction. As shown by 
\citet{nis05} and \citet{dav11} \.{M}$_{\rm out}$ of the molecular gas component can be 10 
to 1000 times lower than \.{M}$_{\rm out}$ from the atomic component.

The ratio of mass outflow to mass accretion, \.{M}$_{\rm out}$/ \.{M}$_{\rm acc}$, is 
typically found to be of the order of 1/10 for Classical T Tauri stars, while there
is a large scatter in the measured ratios for younger YSOs, from 0.01 to 0.9 depending
on tracers used \citep{ant08}, while \citet{car12} find an average ratio of 0.01 for 
Class\,I and very young YSOs when using \.{M}$_{\rm out} (H_2)$. If we can assume that the 
same holds here, we are witnessing a burst of mass accretion about 70 years ago with a 
mass accretion rate \.{M}$_{\rm acc} \sim 10^{-4}$~M$_{\odot}$~yr$^{-1}$.

If the velocities were constant, the \.{M}$_{\rm out}$ found for a knot would provide 
a measure of the instantaneous mass flow rate at a given time in the past, and hence give 
an idea about the accretion history. Although we know from the multi-epoch proper motions 
that the velocities of many $H_2$ emitting knots may vary with time, we take as a first 
approximation the average velocities (over 10 years) at face value, as done also when 
discussing dynamical age. We note the tendency for the space velocities to decrease with
distance from the driving source, in particular for the southern arm of the Class\,0 jet 
(Flow~1). Whether this is due to a slowing down of the jets, which could be expected 
as these shocks travel in a dense medium and may interact with the surrounding gas 
(entrainment), or it reflects a real variation in ejection speed, remains unclear. Also, 
as mentioned above, it is not clear what percentage of the total mass flux is traced by 
the shocked molecular hydrogen. 

The three protostellar flows studied here are located in the same star forming core, 
hence, at the same distance from us, presumably in very similar physical conditions, and 
in addition the three driving sources are of similar bolometric luminosities. The knots 
in the Class\,0 jet are both faster and more massive than the knots in the two Class\,I 
jets, suggesting a clear difference between the two evolutionary stages and supporting 
the idea that accretion and ejection is stronger during the Class\,0 stage, as suggested
by \citet{bont96}. Nevertheless, in view of the emerging picture of the episodic nature of 
the accretion/ejection phenomenon, which is also supported by our time-variable velocities, 
we can not exclude that the observed difference between the Class\,0 and the two Class\,Is 
presented here is also time-variable. Thus, it could be that we by chance study these 
sources at a moment dominated by the effects of a recent massive accretion burst in the 
Class\,0 source.

\section{Summary and Conclusions}
\label{con}

   \begin{enumerate}
      \item We measured proper motions for 31 (out of 57 detected) knots in 
            protostellar jets in the Serpens NH3 region (aka Serpens cluster B), 
            using multi-epoch near-IR $\mathrm{H}_2$ line imaging (2.1218 $\mu$m) 
            with the same telescope/instrument setup over the period 2003 to 2013.

      \item We find tangential velocities over the time-scales: 10 years, 8 years, 
            6 years, 4 years and 2 years for most of these knots. The 10 year 
            base-line proper motions give typical velocities around 50 km/s and a
            maximum velocity of 190 km/s. While these can be viewed as the average 
            velocities over 10 years, the shorter base-lines show evidence of
            time-variable velocities. 

      \item The knot fluxes over the 4 measured epochs are relatively stable and
            flux variability is detected for 45\% of the knots, but only a few vary
            by as much as 30\% or more in their flux. We find no correlation between
            flux variability and velocity variability, nor between flux variability
            and speed. 

      \item There was a burst in brightness around MHO~2235-A1 in 2011 as compared 
            to 2009 with a disappearance again in 2013, a clear piece of evidence 
            of the episodic nature of the jets.

      \item Radial velocities were obtained for the knots for which we have K-band
            spectra, using all the available and sufficiently strong $\mathrm{H}_2$ 
            lines (typically 4-5 lines), yielding the space velocities.
      
      \item Based on the knot velocities, the morphology, and published data on 
            protostars in the region, we interpret our data to show at least three 
            different jet flows in the field we monitored, of which two are bipolar. 
            The three driving sources are suggested to be: MMS3 (Ser-emb-1), a low 
            luminosity Class\,0 source, Ser-emb-11, a binary Class\,I, and Ser-emb-17, 
            another Class\,I source, all deeply embedded and previously studied from 
            mid-IR to longer wavelengths.

      \item The dynamical ages of different parts of these flows are estimated, based 
            on their separations from the driving source and their current velocities. 
            These deduced ages are all $<$ 5000 yrs, the youngest features $<$ 100 yrs.

      \item Time-variable velocities are found in the ejecta from a Class\,0 source,
            for the first time. We also find that the Class\,0 jet has larger measured 
            velocities than the two Class\,I jets.

      \item From the K-band spectra obtained for a subset of the brighter knots, we
            estimate the excitation temperatures, the average value being 
            T$_{\rm ex}$ = 2200 $\pm$ 350 K, which is typical for ro-vibrational states 
            of $\mathrm{H}_2$ in post-shock layers. Towards these knots we have also 
            made a rough estimate of the extinction using the H$_2$ lines, and found
            values as high as A$_V$ = 58 magnitudes. Column densities of H$_2$ were
            obtained from the dereddened line intensities. For 24 knots without spectra
            we estimated column densities through the v=1-0 S(1) line intensity.

     \item Column densities are combined with space velocities and knot location to 
            derive the mass loss rates in the knots. We find that \.{M}$_{out}$ for the 
            Class\,0 jet is about two orders of magnitude larger than \.{M}$_{out}$ for 
            the two Class\,I jets while the L$_{\rm bol}$ of the driving sources are all 
            practically the same.

   \end{enumerate}

\begin{acknowledgements}
We thank the anonymous referee for suggestions that led to a substantial improvement of 
the article. AAD thanks A.J. Delgado and G. Gahm for fruitful discussions.

This work was initiated as one of the student projects during the Nordic-Baltic 
Optical/NIR and Radio Astronomy Summer School held in the Tuorla Observatory, 
Turku, Finland on 8th-18th June 2009, where the financial support from NordForsk 
is greatly acknowledged. 

We have made extensive use of SAOImage DS9, developed by Smithsonian Astrophysical 
Observatory. 

This work made use of EURO-VO software TOPCAT, which is funded by the European 
Commission through contracts RI031675 (DCA) and 011892 (VO-TECH) under the 6th 
Framework Programme and contracts 212104 (AIDA) and 261541 (VO-ICE) under the 7th 
Framework Programme.

TL acknowledges the support of the Estonian Ministry for Education and Science 
as well as European Social Fund's Doctoral Studies and Internationalisation 
Programme DoRa and Kristjan Jaak Scholarship, which are carried out by Foundation 
Archimedes. 

This work is partly based on observations made with the Spitzer Space Telescope, 
which is operated by the Jet Propulsion Laboratory, California Institute of Technology 
under a contract with NASA.

\end{acknowledgements}

\bibliography{serpjet_ref}

\end{document}

%% file: propermotion.tex

\newcommand\cola {\null}
\newcommand\colb {&}
\newcommand\colc {&}
\newcommand\cold {&}
\newcommand\cole {&}
\newcommand\colg {&}
\newcommand\colh {&}
\newcommand\colj {&}
\newcommand\colk {&}
\newcommand\colm {&}
\newcommand\coln {&}
\newcommand\colp {&}
\newcommand\colq {&}
\newcommand\cols {&}
\newcommand\colt {&}
\newcommand\eol{\\}
\newcommand\extline{&&&&&&&&&&&&&&\eol}

\centering
\begin{tabular}{lrrrrrrrrrrrrrr}
\hline\hline
\cola NAME\colb $\alpha_{4}$ (2000)\colc $\delta_{4}$ (2000)\cold PM$_{12}$\cole PA$_{12}$\colg PM$_{23}$\colh PA$_{23}$\colj PM$_{13}$\colk PA$_{13}$\colm PM$_{24}$\coln PA$_{24}$\colp PM$_{34}$\colq PA$_{34}$\cols PM$_{14}$\colt PA$_{14}$\eol

\cola \colb hour\colc deg\cold mas/yr\cole deg\colg mas/yr\colh deg\colj mas/yr\colk deg\colm mas/yr\coln deg\colp mas/yr\colq deg\cols mas/yr\colt deg\eol

\cola MHO\colb $\pm 0.06''$\colc $\pm 0.06''$\cold $\pm$ 10\cole \colg $\pm$ 28\colh \colj $\pm$ 7\colk \colm $\pm$ 15\coln \colp $\pm$ 35\colq \cols $\pm$ 6\colt \eol
\hline
\extline
\cola 2235-M5\colb 18:28:59.296\colc   0:29:13.32\cold   \cole    \colg     92\colh  340$\pm$18\colj   \colk  \colm  \coln  \colp     77\colq  170$\pm$25\cols   \colt  \eol
\cola 2235-M4\colb 18:28:59.666\colc   0:29:12.89\cold     36\cole 350$\pm$16 \colg   \colh \colj     39\colk    4$\pm$11\colm   \coln  \colp   \colq   \cols     22\colt    8$\pm$16\eol
\cola 2235-M1\colb 18:28:59.730\colc   0:29:10.95\cold   \cole  \colg   \colh  \colj   \colk   \colm   \coln   \colp   \colq  \cols   \colt  \eol
\cola 2235-M2\colb 18:28:59.826\colc   0:29:10.24\cold   \cole  \colg   \colh  \colj   \colk  \colm   \coln   \colp   \colq  \cols   \colt  \eol
\cola 2235-I1\colb 18:29:01.722\colc   0:29:24.28\cold   \cole  \colg   \colh  \colj   \colk  \colm   \coln   \colp   \colq  \cols     13\colt 199$\pm$26\eol
\cola 2235-I2\colb 18:29:01.732\colc   0:29:21.97\cold    24\cole 221$\pm$24 \colg   \colh  \colj     23\colk 195$\pm$19\colm     68\coln  168$\pm$13\colp    108\colq  178$\pm$18\cols     37\colt 186$\pm$9\eol
\cola 2235-I3\colb 18:29:01.746\colc   0:29:26.29\cold    23\cole  324$\pm$25 \colg     69\colh   89$\pm$23\colj     16\colk   32$\pm$ 26\colm   \coln \colp   \colq \cols     13\colt  341$\pm$27\eol
\cola 2235-K\colb 18:29:01.953\colc   0:29:04.37\cold     68\cole  306$\pm$9 \colg   \colh  \colj     41\colk  300$\pm$ 11\colm     61\coln  158$\pm$15\colp     91\colq  164$\pm$21\cols     24\colt 273$\pm$15\eol
\cola 2235-F2\colb 18:29:03.676\colc   0:29:56.20\cold    53\cole  194$\pm$11\colg    128\colh  269$\pm$13\colj     58\colk 29$\pm$ 7\colm     42\coln 251$\pm$21\colp     74\colq  113$\pm$26\cols     43\colt 213$\pm$8\eol
\cola 2235-F1\colb 18:29:03.843\colc   0:29:59.87\cold   \cole  \colg   \colh  \colj   \colk   \colm   \coln   \colp     79\colq  179$\pm$  24\cols   \colt  \eol
\cola 2235-E\colb 18:29:04.928\colc   0:30:15.12\cold     91\cole 235$\pm$6 \colg    113\colh  164$\pm$14\colj     82\colk 214$\pm$5\colm    113 \coln 198$\pm$8\colp    158\colq 228$\pm$13\cols     95\colt 218$\pm$3\eol
\cola 2235-D\colb 18:29:05.406\colc   0:30:21.11\cold     24\cole  162$\pm$24 \colg     89\colh   36$\pm$18\colj     19\colk   83$\pm$22\colm   \coln  \colp     78\colq  178$\pm$25\cols     20\colt  127$\pm$17\eol
\cola 2235-C1\colb 18:29:05.997\colc   0:30:22.06\cold    48\cole  230$\pm$12 \colg   \colh \colj     43\colk 242$\pm$10\colm   \coln   \colp   \colq   \cols     28\colt 235$\pm$13\eol
\cola 2235-A1\colb 18:29:06.655\colc   0:30:30.95\cold     34\cole  170$\pm$17 \colg   \colh  \colj     28\colk  167$\pm$15\colm   \coln   \colp   \colq   \cols     27\colt  167$\pm$13\eol
\cola 2235-A2\colb 18:29:06.761\colc   0:30:34.05\cold   \cole  \colg   \colh  \colj   \colk  \colm   \coln   \colp  \colq   \cols   \colt  \eol
\cola 2235-B\colb 18:29:06.934\colc   0:30:24.77\cold   \cole  \colg   \colh  \colj     15\colk  146$\pm$ 27\colm   \coln   \colp   \colq   \cols     13\colt  133$\pm$26\eol
\cola 2234-H8\colb 18:29:07.922\colc   0:30:24.86\cold     33\cole 312$\pm$18\colg     67\colh   88$\pm$24\colj     17\colk    0$\pm$25\colm   \coln  \colp   \colq   \cols     18\colt 334$\pm$ 20\eol
\cola 2234-H3\colb 18:29:08.153\colc   0:30:31.95\cold     23\cole 212$\pm$25\colg     61\colh  136$\pm$26\colj     26\colk  175$\pm$16\colm   \coln  \colp   \colq   \cols     13\colt  167$\pm$26\eol
\cola 2234-G1\colb 18:29:08.235\colc   0:30:47.98\cold     41\cole 207$\pm$14\colg    130\colh  350$\pm$13\colj     21\colk  291$\pm$20\colm   \coln  \colp    119\colq  169$\pm$17\cols     19\colt 219$\pm$19\eol
\cola 2234-H1\colb 18:29:08.363\colc   0:30:32.56\cold     51\cole 194$\pm$12 \colg   \colh  \colj     42\colk 186$\pm$10\colm   \coln   \colp  \colq   \cols     36\colt 188$\pm$10\eol
\cola 2234-H5\colb 18:29:08.469\colc   0:30:39.93\cold   \cole  \colg   \colh  \colj   \colk   \colm   \coln   \colp  \colq   \cols   \colt   \eol
\cola 2234-H2\colb 18:29:08.498\colc   0:30:35.59\cold   \cole  \colg   \colh  \colj     16\colk  149$\pm$26\colm   \coln   \colp  \colq  \cols   \colt  \eol
\cola 2234-C\colb 18:29:08.609\colc   0:31:13.50\cold     27\cole    5$\pm$22 \colg    125\colh 189$\pm$13\colj   \colk  \colm     34\coln 186$\pm$25\colp     76\colq   12$\pm$25\cols   \colt  \eol
\cola 2234-E\colb 18:29:08.610\colc   0:31:07.55\cold     23\cole 200$\pm$25\colg    106\colh 207$\pm$15\colj     45\colk 204$\pm$10\colm   \coln  \colp    144\colq   18$\pm$14\cols   \colt  \eol
\cola 2234-H4\colb 18:29:08.610\colc   0:30:38.89\cold    38\cole 214$\pm$16\colg   67\colh   90$\pm$24\colj     23\colk  175$\pm$19\colm     39\coln   92$\pm$23\colp   \colq \cols     19\colt  173$\pm$18\eol
\cola 2234-D1+2\colb 18:29:08.810\colc   0:31:08.38\cold  74\cole 168$\pm$8\colg    88\colh  183$\pm$18\colj     77\colk  173$\pm$5\colm   \coln  \colp   \colq   \cols     56\colt  167$\pm$6\eol
\cola 2234-B1\colb 18:29:09.009\colc   0:31:22.08\cold    52\cole 179$\pm$11\colg   149\colh 203$\pm$11\colj     76\colk  191$\pm$6\colm     70\coln 198$\pm$13\colp   \colq \cols     58\colt 188$\pm$6\eol
\cola 2234-A1\colb 18:29:09.045\colc   0:31:28.10\cold  \cole \colg  \colh \colj  \colk  \colm   \coln \colp   \colq \cols   \colt \eol
\cola 2234-A2\colb 18:29:09.057\colc   0:31:26.64\cold    71\cole 180$\pm$8\colg    116\colh 197$\pm$14\colj     82\colk  186$\pm$5\colm     46\coln 190$\pm$19\colp   \colq \cols     61\colt 183$\pm$6\eol
\cola 2234-I1\colb 18:29:09.570\colc   0:32:07.53\cold   \cole  \colg   \colh  \colj   \colk   \colm     37\coln   35$\pm$24\colp     70\colq   54$\pm$27\cols     19\colt   63$\pm$19\eol
\cola 2233-N\colb 18:29:11.560\colc   0:31:20.05\cold     23\cole  100$\pm$25\colg   \colh  \colj   \colk   \colm     36\coln   14$\pm$24\colp   \colq \cols     21\colt   56$\pm$17\eol
\cola 2233-P\colb 18:29:12.443\colc   0:31:46.19\cold     23\cole  125$\pm$25\colg   \colh  \colj   \colk  \colm     57\coln  316$\pm$16\colp   \colq  \cols   \colt   \eol
\hline
\end{tabular}


%% file: fluxes2.tex

\newcommand\cola {\null}
\newcommand\colb {&}
\newcommand\colc {&}
\newcommand\cold {&}
\newcommand\cole {&}
\newcommand\colf {&}
\newcommand\colg {&}
\newcommand\eol{\\}
\newcommand\extline{&&&&&\eol}

\begin{tabular}{lrrrrrr}
\hline
\cola NAME\colb f$_1$\colc f$_2$\cold f$_3$\cole f$_4$\colf  R$_{ap}$\colg Var \eol
\cola MHO\colb \colc \cold \cole \colf   ('')\colg \eol
\hline
\hline
\cola 2233-N \colb  6.47\colc  6.77\cold    \cole  6.30\colf    1.87\colg y\eol
\cola 2233-P\colb  1.41\colc  1.51\cold    \cole  1.54\colf    1.40\colg n\eol
\hline
\cola 2234-A1+A2\colb  4.75\colc  4.48\cold  4.15\cole  4.25\colf    1.17\colg y\eol
\cola 2234-B1\colb  5.67\colc  5.74\cold  5.32\cole  5.70\colf    1.64\colg n\eol
\cola 2234-C\colb  7.24\colc  7.73\cold  7.39\cole  7.63\colf    2.81\colg y\eol
\cola 2234-D1+D2\colb 21.06\colc 20.08\cold 20.10\cole 20.60\colf    3.28\colg y\eol
\cola 2234-E\colb  6.25\colc  7.33\cold  8.20\cole  8.48\colf    2.34\colg y\eol
\cola 2234-G1\colb  1.03\colc  1.35\cold  1.45\cole  1.22\colf    1.87\colg n\eol
\cola 2234-H1\colb 22.55\colc 23.25\cold 23.94\cole 22.61\colf    2.57\colg y\eol
\cola 2234-H2\colb  6.97\colc  6.89\cold  7.12\cole  6.70\colf    1.87\colg n\eol
\cola 2234-H3\colb  3.89\colc  3.98\cold  4.08\cole  3.96\colf    1.40\colg n\eol
\cola 2234-H4\colb  1.83\colc  1.54\cold  1.81\cole  1.86\colf    1.17\colg n\eol
\cola 2234-H5\colb  2.34\colc  2.29\cold  2.40\cole  2.20\colf    1.40\colg n\eol
\cola 2234-H8\colb  2.54\colc  3.01\cold  3.17\cole  2.73\colf    2.34\colg y\eol
\hline
\cola 2235-A1\colb 11.86\colc 12.05\cold 14.56\cole 11.33\colf    2.34\colg y\eol
\cola 2235-A2\colb  2.40\colc  2.40\cold  2.93\cole  3.06\colf    1.40\colg y\eol
\cola 2235-B\colb 48.08\colc 46.13\cold 48.30\cole 44.91\colf    3.74\colg y\eol
\cola 2235-C1\colb  3.67\colc  3.00\cold  3.23\cole  3.06\colf    1.87\colg y\eol
\cola 2235-D\colb  1.57\colc  1.31\cold  1.13\cole  1.66\colf    1.87\colg y\eol
\cola 2235-E\colb  6.24\colc  6.50\cold  7.78\cole  6.64\colf    7.02\colg y\eol
\cola 2235-F1\colb  2.41\colc  2.40\cold  2.34\cole  2.42\colf    1.64\colg n\eol
\cola 2235-I1\colb  0.96\colc  1.05\cold  1.14\cole  1.25\colf    1.17\colg n\eol
\cola 2235-I2\colb  0.96\colc  1.21\cold  1.11\cole  1.29\colf    1.40\colg n\eol
\cola 2235-I3\colb  0.60\colc  0.59\cold  0.50\cole  0.57\colf    1.17\colg n\eol
\cola 2235-K\colb  \colc  2.46\cold  2.38\cole  2.61\colf    1.64\colg n\eol
\cola 2235-M1\colb    \colc  4.46\cold  4.33\cole  4.02\colf    1.17\colg n\eol
\cola 2235-M2\colb    \colc  3.37\cold  3.28\cole  2.93\colf    1.40\colg n\eol
\cola 2235-M4\colb     \colc  4.20\cold  4.00\cole  3.90\colf    1.17\colg n\eol
\cola 2235-M5\colb    \colc  0.93\cold  0.93\cole  1.01\colf    1.17\colg n\eol
\hline
\end{tabular}


%% file: mass_ordered2.tex

\newcommand\cola {\null}
\newcommand\colb {&}
\newcommand\colc {&}
\newcommand\cold {&}
\newcommand\cole {&}
\newcommand\eol{\\}
\newcommand\extline{&&&&\eol}

\begin{tabular}{lcccr}
\hline
\extline
\cola Knot\colb    N$_{H_2}$\colc      M$_{H_2}$\cold \.M$_{\rm out}(H_2)$\cole  n$_{H_2}$\eol
\cola MHO  \colb   (cm$^{-2}$)\colc (M$_{\odot}$)\cold     (M$_{\odot}$ yr$^{-1}$)\cole (cm$^{-3}$)\eol
\extline
\hline
\hline

\hline
\cola 2234-A1+A2$^{*}$\colb 1.5 $\times 10^{20}$\colc 7.0 $\times 10^{-5}$\cold 9.6 $\times 10^{-7}$\cole 16900\eol
\cola         \colb (2.8 $\times 10^{19}$)\colc (1.3 $\times 10^{-5}$)\cold (1.9 $\times 10^{-7}$)\cole (3200)\eol
\cola 2234-B1\colb 2.2 $\times 10^{19}$\colc 7.0 $\times 10^{-6}$\cold 4.7 $\times 10^{-8}$\cole    2500\eol
\cola 2234-C\colb 3.7 $\times 10^{18}$\colc 1.1 $\times 10^{-6}$\cold 1.6 $\times 10^{-9}$\cole      4300\eol
\cola 2234-D1+D2$^{*}$\colb 1.4 $\times 10^{18}$\colc 6.4 $\times 10^{-7}$\cold 1.5 $\times 10^{-9}$\cole  200\eol
\cola        \colb (1.2 $\times 10^{18}$)\colc (5.6 $\times 10^{-7}$)\cold (1.4 $\times 10^{-9}$)\cole (100)\eol

\cola 2234-E\colb 5.2 $\times 10^{17}$\colc 2.1 $\times 10^{-7}$\cold 3.9 $\times 10^{-10}$\cole     60\eol
\cola 2234-G1\colb 7.7 $\times 10^{17}$\colc 6.6 $\times 10^{-8}$\cold 2.8 $\times 10^{-11}$\cole    90\eol
\cola 2234-H1$^{*}$\colb 1.1 $\times 10^{19}$\colc 5.0 $\times 10^{-6}$\cold 3.0 $\times 10^{-9}$\cole  1300\eol
\cola       \colb (8.4 $\times 10^{18}$)\colc (3.9 $\times 10^{-6}$)\cold (2.4 $\times 10^{-9}$)\cole (1000)\eol

\cola 2234-H2\colb 3.6 $\times 10^{18}$\colc 1.0 $\times 10^{-6}$\cold 3.0 $\times 10^{-10}$\cole   400\eol
\cola 2234-H3\colb 2.4 $\times 10^{18}$\colc 6.0 $\times 10^{-7}$\cold 2.6 $\times 10^{-10}$\cole   300\eol
\cola 2234-H4\colb 1.4 $\times 10^{18}$\colc 2.6 $\times 10^{-7}$\cold 9.7 $\times 10^{-11}$\cole   200\eol
\cola 2234-H5\colb 2.3 $\times 10^{18}$\colc 3.5 $\times 10^{-7}$\cold 3.4 $\times 10^{-11}$\cole   300\eol
\cola 2234-H8\colb 2.1 $\times 10^{18}$\colc 4.3 $\times 10^{-7}$\cold 1.1 $\times 10^{-10}$\cole   300\eol
\hline
\cola 2233-N$^{*}$\colb 4.4 $\times 10^{17}$\colc 2.3 $\times 10^{-7}$\cold 5.7 $\times 10^{-11}$\cole  50\eol 
\cola   \colb (1.6 $\times 10^{17})$\colc (8.5 $\times 10^{-8}$)\cold (2.1 $\times 10^{-11})$\cole (20)\eol

\cola 2233-P\colb 9.2 $\times 10^{16}$\colc 2.0 $\times 10^{-8}$\cold 4.0 $\times 10^{-12}$\cole    10\eol
\cola 2235-A2\colb 9.9 $\times 10^{17}$\colc 1.4 $\times 10^{-7}$\cold 8.1 $\times 10^{-11}$\cole  100\eol
\cola 2235-C1\colb 1.1 $\times 10^{18}$\colc 1.7 $\times 10^{-7}$\cold 3.0 $\times 10^{-10}$\cole  100\eol
\cola 2235-D\colb 6.1 $\times 10^{17}$\colc 7.4 $\times 10^{-8}$\cold 6.5 $\times 10^{-11}$\cole    70\eol
\cola 2235-E\colb 1.4 $\times 10^{18}$\colc 3.5 $\times 10^{-7}$\cold 1.0 $\times 10^{-9}$\cole    200\eol
\cola 2235-F1\colb 9.0 $\times 10^{17}$\colc 1.2 $\times 10^{-7}$\cold 1.8 $\times 10^{-10}$\cole  100\eol
\cola 2235-I1\colb 5.2 $\times 10^{17}$\colc 5.7 $\times 10^{-8}$\cold 7.4 $\times 10^{-12}$\cole   60\eol
\cola 2235-I2\colb 5.6 $\times 10^{17}$\colc 5.9 $\times 10^{-8}$\cold 2.2 $\times 10^{-11}$\cole   60\eol
\cola 2235-I3\colb 3.0 $\times 10^{17}$\colc 2.9 $\times 10^{-8}$\cold 6.7 $\times 10^{-12}$\cole   30\eol
\cola 2235-K\colb 8.1 $\times 10^{17}$\colc 1.3 $\times 10^{-7}$\cold 2.7 $\times 10^{-11}$\cole    90\eol
\cola 2235-M1\colb 8.5 $\times 10^{17}$\colc 2.2 $\times 10^{-7}$\cold 5.1 $\times 10^{-12}$\cole   100\eol
\cola 2235-M2\colb 7.3 $\times 10^{17}$\colc 1.7 $\times 10^{-7}$\cold 1.1 $\times 10^{-11}$\cole   80\eol
\cola 2235-M4\colb 7.4 $\times 10^{17}$\colc 2.1 $\times 10^{-7}$\cold 3.6 $\times 10^{-11}$\cole   90\eol
\cola 2235-M5\colb 4.9$\times 10^{17}$\colc 5.0 $\times 10^{-8}$\cold 2.8 $\times 10^{-11}$\cole    60\eol
\hline
\cola 2235-A1\colb 1.3 $\times 10^{18}$\colc 6.5 $\times 10^{-7}$\cold 1.2 $\times 10^{-9}$\cole   200\eol
\cola 2235-B$^{*}$\colb 1.1 $\times 10^{19}$\colc 8.0 $\times 10^{-6}$\cold 4.8 $\times 10^{-9}$\cole 1300\eol
\cola     \colb (4.8 $\times 10^{18}$)\colc (3.4 $\times 10^{-6}$)\cold (2.1 $\times 10^{-9}$)\cole (600)\eol

\hline

\end{tabular}

$^{*}$ Measurements using the spectroscopic data. In parenthesis below are
given the values obtained on the same knot using the 2.1218 $\mu$m imaging
method.
